\documentclass{article}
\usepackage[utf8]{inputenc}
\usepackage{amsmath}
\usepackage{amssymb}
\usepackage{bbm}
\usepackage{setspace,graphicx,amssymb,amsmath,latexsym,amsfonts,amscd,amsthm,multirow,ctable,mathdots,caption,array,diagbox,mathtools}
\usepackage{color}
\usepackage{subcaption}
\usepackage{jheppub}

\newcommand{\scO}{\mathcal{O}}
\newcommand{\CO}{\mathcal{O}}

\newcommand{\CS}{\mathcal{S}}
\newcommand{\CT}{\mathcal{T}}
\newcommand{\<}{\langle}
\renewcommand{\>}{\rangle}
\newcommand{\pd}{\partial}

\title{Bootstrapping the gap in quantum spin systems}
\author{Colin Oscar Nancarrow$^1$, }
\affiliation{$^1$Department of Physics, Boston University, 
Boston, MA  02215, USA}
\author{Yuan Xin$^{2}$}
\affiliation{$^2$Department of Physics, Yale University, New Haven, CT 06520, USA
}
\abstract{
    In this work we report on a new bootstrap method for quantum mechanical problems that closely mirrors the setup from conformal field theory (CFT). We use the equations of motion to develop an analogue of the conformal block expansion for matrix elements and impose crossing symmetry in order to place bounds on their values. The method can be applied to any quantum mechanical system with a local Hamiltonian, and we test it on an anharmonic oscillator model as well as the $(1+1)$-dimensional transverse field Ising model (TFIM). For the anharmonic oscillator model we show that a small number of crossing equations provides an accurate solution to the spectrum and matrix elements. For the TFIM we show that the Hamiltonian equations of motion, translational invariance and global symmetry selection rules imposes a rigorous bound on the gap and the matrix elements of TFIM in the thermodynamic limit. The bound improves as we consider larger systems of crossing equations, ruling out more finite-volume solutions. 
    Our method provides a way to probe the low energy spectrum of an infinite lattice from the Hamiltonian rigorously and without approximation.
}

\begin{document}

\maketitle


\section{Introduction}

Determining the macroscopic phase of matter from the microscopic description of fundamental interactions is a central topic in physics. The problem is challenging as simple microscopic models can become strongly coupled at large scale, and perturbation expansion breaks down. Monte-Carlo simulation and variational methods are useful non-perturbative tools. Both methods are enormously successful, but still have limitations. Monte-Carlo simulation has statistical errors. Finite-size effects can obscure the symmetry at the critical point, and sign problems remain an obstacle in many systems. Variational methods are usually exponentially hard to converge to the thermodynamic limit, and special methods like MPS that have power-law convergence only demonstrate superiority in low-dimensional systems. The bootstrap philosophy in physics has recently undergone a renaissance which has produced strong bounds on the dynamical data in conformal field theories\cite{Rattazzi:2008pe,El-Showk:2012cjh,Poland:2018epd}. The bootstrap approach is different from the first principle calculation approach described above as bootstrap does not rely on a microscopic description. This gives bootstrap the power to place rigorous, generic bounds on dynamical data but also limits its application when we wish to implement more knowledge of the microscopic description. More recently, a new bootstrap paradigm that leverages the unitarity and microscopic equations of motion emerged. 
The new method produces rigorous bounds using the explicit microscopic Lagrangian and Hamiltonian, making it a promising tool to study strongly coupled systems from first principles. The method was first applied to large $N$ lattice Yang-Mills theory \cite{Anderson:2016rcw,anderson2018loop} and was followed up by \cite{Kazakov:2022xuh} with a more careful treatment of symmetries and a more efficient algorithm. These works consider a large number of Wilson loop expectation values and show that the space of such expectation values are constrained by reflection positivity and Schwinger-Dyson equations of motion. The method was also used to study the lattice Ising model\cite{Cho:2022lcj} and matrix models \cite{Lin:2020mme, Kazakov:2021lel}. For hamiltonian systems there is a similar method. Positivity of the Hilbert space inner product imposes a positive semidefinite constraint on the moment matrix of eigenstates. The equations of motion are used to reduce the number of independent parameters in this matrix and the resulting parameter space is partitioned by the constraint. The method has been applied to various quantum mechanical \cite{Han:2020bkb,Berenstein:2021dyf,Bhattacharya:2021btd,Berenstein:2021loy,Berenstein:2022ygg,Berenstein:2022unr,Morita:2022zuy} and lattice models\cite{han2020quantum,Lawrence:2021msm,Blacker_2022}. Similar methods also exist for classical mechanical systems \cite{doi:10.1137/15M1053347,Nakayama:2022ahr} and non-unitary systems\cite{Li:2022vzn,Khan:2022uyz}. 

These lattice and moment matrix bootstrap methods are successful both conceptually as the bound is rigorous and numerically because the bounds converge rapidly when the system has small correlation length, but there is still room for improvement. The infrared (IR) resolution of lattice bootstrap is limited by the spacetime span of the operators studied, which is naively exponentially difficult to increase. The moment matrix bootstrap setup only constrains the observables on the same in- and out-states thus many dynamical observables are unbounded. Moreover, no existing bootstrap method produces bounds on the gap of a lattice theory. Therefore, we are motivated to look for more constraints that have not been considered by current methods. A possible generalization is by systematically introducing observables that are off-diagonal in terms of the in- and out- states. However, as we will show in the next section, the straightforward generalization fails to improve the bound and a more sophisticated approach needed.

\begin{figure}
    \centering
    \includegraphics[width=0.6\linewidth]{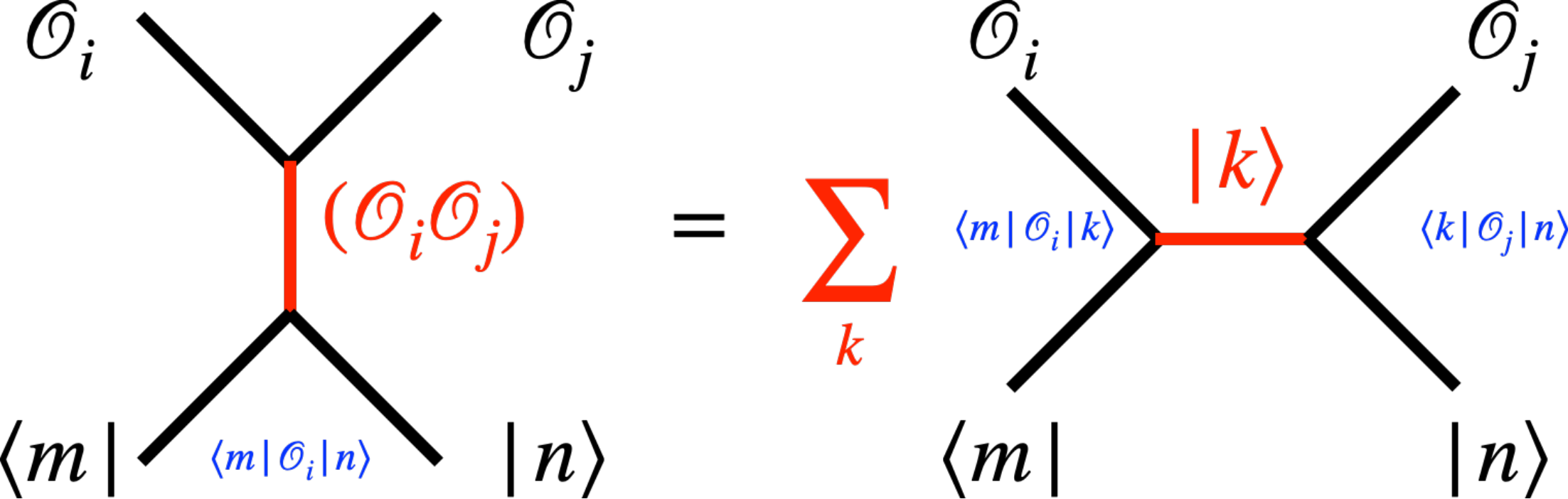}
    \caption{A schematic demonstration of the spectral bootstrap crossing equation. The lhs and rhs are analogous to the t-channel and s-channel, respectively, that arise as a consequence of OPE associativity in the conformal bootstrap. In the t-channel, we contract the operator $\CO_i$ with $\CO_j$ and obtain a third operator. In the s-channel, we contract the operators with the external states, and sum over a complete set of exchanged states.   }
    \label{fig:schematica-crossing}
\end{figure}
In this work we report on a new bootstrap method called the {\it spectral bootstrap} that closely mirrors the structure of the conformal bootstrap. We consider the spectral decomposition of any two-point correlator on a pair of energy eigenstates
\begin{equation}\label{eq:crossing-eq}
    \< i | \CO_1 \CO_2 | j \> = \sum_k \< i | \CO_1 | k \> \< k | \CO_2 | j \>
\end{equation}
to be our ``crossing equations''. A schematic demonstration of the crossing equation is shown in Figure~\ref{fig:schematica-crossing}. These equations can be organized in terms of quadratic forms of the one-point matrix elements $\< i | \CO | k \>$ which are analogous to the ``OPE coefficients''. The space of independent one-point matrix elements can be systematically reduced to ``primary elements'' using the Hamiltonian equations of motion, reminiscent of the conformal block decomposition. We use semidefinite programming to exploit the constraints on the positivity of the quantities \(\left| \< i | \scO | k \> \right|^2\). Table~\ref{tab:analogy} contains the dictionary between objects appearing in the conformal and spectral bootstraps. The spectral bootstrap manifestly contains the unitarity constraints imposed by the moment matrix approach as a special case where $ \CO_1$ and $\langle i |$ are conjugate to $\CO_2$ and $|j \rangle$, respectively. The spectral bootstrap is stronger as it imposes constraints on diagonal and off-diagonal matrix elements jointly. More observables such as the transition rate and the gap become accessible in the spectral bootstrap and the existing bounds from the matrix approach are expected to become stronger. 

We test the spectral bootstrap method on an anharmonic oscillator model and the $(1+1)D$ transverse field ising model (TFIM). As a warm-up, we show that the bounds on the low-lying energy eigenvalues $E_0$ and $E_1$ of a double-well potential are drastically improved by the spectral density compared with the moment matrix approach. We begin by taking the external states $| i \rangle$ and $| j \rangle$ in (\ref{eq:crossing-eq}) to both be the ground state $|0\rangle$, obtaining a setup that is analogous to the single correlator setup in the conformal bootstrap. The spectral bootstrap bounds $E_0$ on a narrow interval and $E_1$ from above. Even at low truncation, the allowed interval of $E_0$ is parametrically smaller than what the moment matrix bootstrap produces at a higher truncation. The $E_1$ upper bound is almost perfectly saturated by the exact solution. We then extend the setup to the mixed matrix elements between the ground state $| 0 \rangle$ and the first excited state $| 1 \rangle$ and find even for minimal truncation all parameters, $E_0$, $E_1$, $\<0|x^2|0\>$, $\<1|x^2|1\>$ and $\<0|x|1\>$ are bounded in a tiny island of precision $\sim 10^{-8}$. We apply the extremal functional method as in \cite{El-Showk:2012vjm,Simmons-Duffin:2016wlq} to extract the spectrum and find the eigenvalues of many higher excited states are in good agreement with the exact result. With TFIM, we show that the spectral decomposition of correlators of the form $\<0|\CO \CO'|0\>$, Hamiltonian equations of motion, translation invariance and spin, parity and time-reversal symmetry together impose non-trivial upper bound on the gap and local operator matrix elements of the infinite lattice model. The bound improves and becomes closer to the exact infinite volume values as we include larger number of equations of motion and crossing equations. 

The paper is organized as follows: In section~\ref{sec:preview} we summarize the key ingredients in the spectral bootstrap method and make an analogy with the conformal bootstrap. In section~\ref{sec:matrix-bootstrap} we study the anharmonic oscillator as a warm-up. We begin by revisiting the matrix bootstrap method and discussing the motivation to switch to the spectral bootstrap approach. Then we introduce the spectral bootstrap and present improved numerical results. In section~\ref{sec:tfim} we study the transverse field Ising model. First we explain the subtleties of the infinite lattice and how to modify the bootstrap setup to address them. Finally we present the numerical results on the TFIM.




\begin{table}[ht]
    \centering
    \begin{tabular}{c|c}
        \hline
        Conformal bootstrap analogy & Spectral decomposition bootstrap objects \\
        \hline
        OPE coefficients & Matrix elements $\< i | \CO_k | j \>$ \\
        Crossing equations & Spectral decomposition $\< i | \CO_1 \CO_2 | j \> = \sum_k \< i | \CO_1 | k \> \< k | \CO_2 | j \>$ \\
        Conformal symmetry & $\begin{cases}
            \text{Hamiltonian equations of motion } \< i | [H,\CO] | j \> = (E_i - E_j)\< i | \CO | j \> \\ 
            \text{Translation invariance } \< i | U^{n}\CO U^{-n} | j \> = e^{i (p_i-p_j)n}\< i | \CO | j \>
        \end{cases}$ \\
        Scaling Dimension $\Delta$ & Energy $(E-E_{\rm vacuum})$ and momentum $p$ \\
        Spin, parity, global symmetry irreps & (discrete) rotation irreps, parity, global symmetry irreps \\
        Descendants $\partial_{\mu_1}\cdots \partial_{\mu_n} (\partial^2)^m\phi$ & $\left. \left[H,\left[H,\cdots \left[H, \CO(x_n) \right]\cdots\right]\right]\right|_{x_n = n\vec{e} }$ \\
        Single-correlator bootstrap & $\< \Omega | \CO_1 \CO_2 | \Omega \>$ where $\CO_1$, $\CO_2$ are descendants of the same operator \\
        Mixed-correlator bootstrap & $\< \Omega | \CO_1 \CO_2 | \Omega \>$, $\< \Omega | \CO_1 \CO_2 | k \>$, $\< k | \CO_1 \CO_2 | k \>$ with generic $\CO_1$, $\CO_2$  \\
        \hline
    \end{tabular}
    \caption{Dictionary of the analogy between the spectral decomposition bootstrap and the conformal bootstrap.}
    \label{tab:analogy}
\end{table}

\subsection{Summary of setup} \label{sec:preview}

In the spectral bootstrap method, the key ingredient is the crossing equation obtained by inserting the identity $\mathbbm{1}=\sum_k |k\> \< k|$ between a product of two operators
\begin{equation}\tag{\ref{eq:crossing-eq}}
    \< i | \CO_1 \CO_2 | j \> = \sum_k \< i | \CO_1 | k \> \< k | \CO_2 | j \> \, ,
\end{equation}
where the external states $| j \>$ are taken to be the energy eigenstates of the Hamiltonian $H$. The crossing equation is analogous to the crossing equations in the conformal bootstrap. The matrix elements $\< i | \CO | j \>$ are analogous to the OPE coefficients of the conformal bootstrap. Whereas in the conformal bootstrap the OPE coefficients can be reduced by conformal symmetry, in the spectral bootstrap we can reduce the number of independent matrix elements using the Hamiltonian equations of motion, translational invariance and global symmetries
\begin{equation}\label{eq:eom}
    \begin{cases}
        \langle n | [H, \scO] | m \rangle = (E_n -E_m)\langle n | \scO | m \rangle \\
        \langle n | U^x\scO U^{-x} | m \rangle = e^{i(p_n - p_m)x} \langle n | \scO | m \rangle \\
        \langle n | \scO | m \rangle = \langle Q n | Q \scO Q^{-1} | Q m \rangle  \, ,
    \end{cases}
\end{equation}
where $U$ denotes the lattice translation operator and $Q$ is a global symmetry generator. In few-body quantum mechanics problems we have additional equations 
\begin{equation}
    \langle n | H \scO | m \rangle = E_n \langle n | \scO | m \rangle \, .
\end{equation}
After imposing the relations (\ref{eq:eom}), the independent matrix elements are reduced to a small basis. To make an analogy with the conformal bootstrap, we define a basis of {\it primary operators} $\mathcal{A} = \{ \CO_{b_1}, \CO_{b_2}, \cdots \}$, and all matrix element are fixed up to the matrix elements of the primary operators
\begin{equation}
    \langle n | \scO | m \rangle = \sum_{i} g^{\scO}_{i}(E_n,E_m, \cdots) \langle n | \scO_{b_i} | m\rangle.
\end{equation}
Thus a generic operator can be taken as the descendant of the primary operators $\scO_{b_i}$. The factors $g^{\scO}_{i}(E_n,E_m, \cdots)$ are known functions that depend on the energy, momentum and global symmetry charges of the external states. If we consider the crossing equations (\ref{eq:crossing-eq}) for all combinations of a set of external states, e.g. $\< m | \CO \CO' | m\>$, $\< m | \CO \CO' | n\>$ and $\< n | \CO \CO' | n\>$ for a pair of external state $|m \>$, $|n \>$, we can summarize all equations in block vector forms
\begin{equation}
    0 = \sum_{n} \vec{c}^T \vec{\mathcal{V}} (E_n,p_n,\cdots) \vec{c} + (\text{ 
other global symmetry sectors...} \,) \, ,
\end{equation}
where the vector $\vec{c}$ contains all the independent primary operator matrix elements $\langle n | \scO_{b_i} | m\rangle$. The block vector $\vec{\mathcal{V}}$ is a vector of $({\rm dim}~\vec{c}\times {\rm dim}~\vec{c})$-matrices, and $\vec{\mathcal{V}}$ contains the energy factors $g^{\scO}_{i}(E_n,E_m, \cdots)$. In the conformal bootstrap, $\vec{\mathcal{V}}$ contains the conformal blocks. 
To impose constraints, we prove by contradiction:
\begin{equation}\label{eq:bootstrap-problem-generic}
\boxed{
\begin{aligned}
&\text{If there exists $\alpha_{I}$ such that $\forall E_{n\neq 0} \geqslant E_{\rm gap}$}\\ 
& \sum_{I} \alpha_{I} \big(\vec{\mathcal{V}}(E_n,p_n)\big)_{I} \succ 0 \\
& \text{then all spectra with the prescribed ($E_{\rm gap}$) are ruled out.}
\end{aligned}
}
\end{equation}
We use semidefinite programming to solve the following problem.

\section{Anharmonic oscillator warm-up}
The quartic potential anharmonic oscillator is a touchstone model in the recent matrix positivity methods for quantum mechanics bootstrap. We will use it as an opportunity to introduce our method.
The Hamiltonian is
\begin{equation}
    H =  p^2 + \omega^2 x^2 + g x^4 \, .
\end{equation}
The perturbative expansion in $g$ does not converge and in the large $g$ regime fails to provide a good approximation to the exact spectrum. 
Instead one can evaluate the Hamiltonian on the harmonic oscillator basis as an infinite dimensional matrix and truncate to obtain an approximate spectrum which converges exponentially to the exact spectrum. It is convenient to take the truncation to be large enough so that the error can be ignored and treat the approximate spectrum as exact.
We will begin by revisiting the matrix bootstrap method, comparing it with exact results. We then introduce the spectral bootstrap method and discuss the advantages. 

\subsection{Matrix bootstrap revisited}\label{sec:matrix-bootstrap}
\begin{figure}
    \centering
    \includegraphics[width=0.6\linewidth]{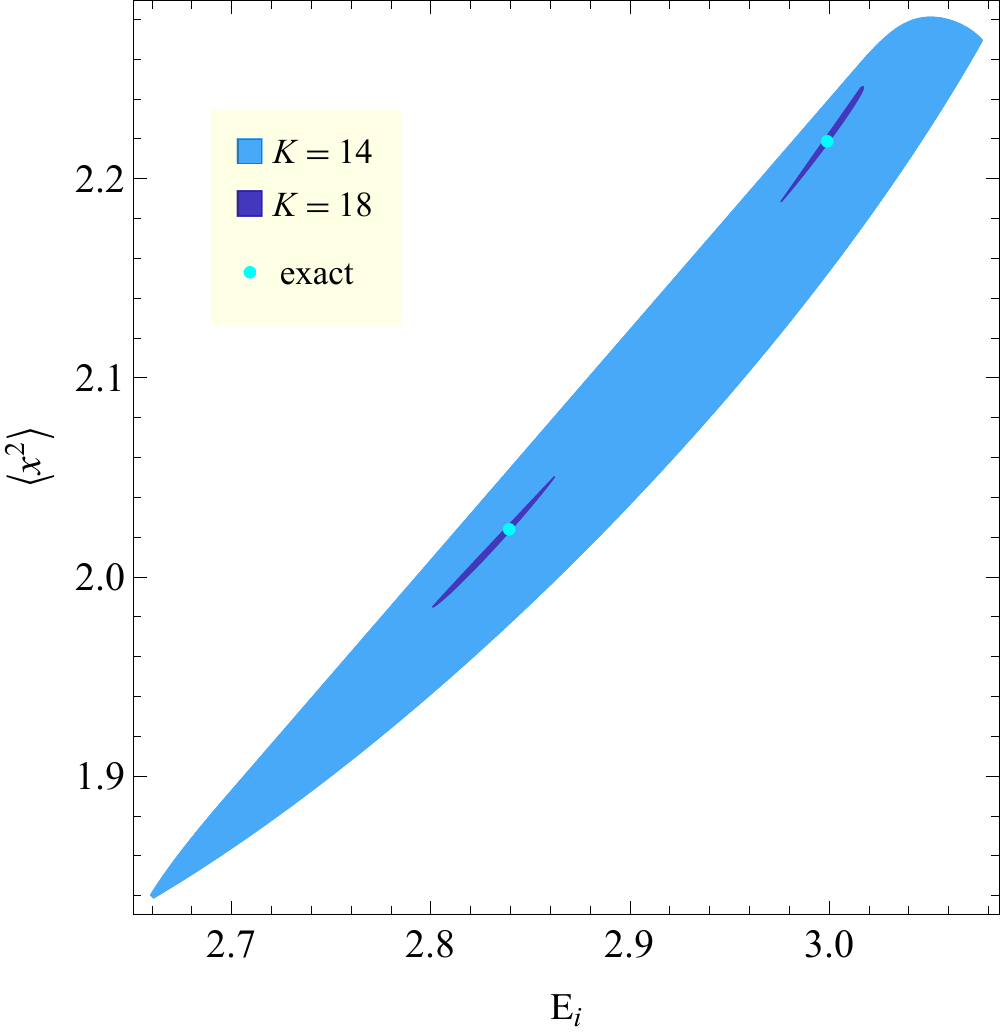}
    \caption{Bound on anharmonic oscillator's energy eigenvalues $E_i$ and $\< x^2 \>$ expectation value from matrix bootstrap (\ref{eq:positivityAHO}). The Hamiltonian is $H = p^2 - 5 x^2 + g^4$. The colored regions represent the allowed parameter space at different truncation $K$ on the Hankel matrix and the cyan points represent the exact eigenvalues and expectation values. The exact solution shows that the lowest two eigenstates are in the plot region. At $K = 14$ the bound does not distinguish the two eigenstates. At $K=18$ the bound is improved and the allowed region splits into two distinct islands surrounding the exact answer.  }
    \label{fig:matrix-bootstrap}
\end{figure}

The anharmonic oscillator has been studied by the matrix bootstrap method\cite{Han:2020bkb,Berenstein:2021dyf,Bhattacharya:2021btd,Berenstein:2021loy,Berenstein:2022ygg,Berenstein:2022unr}. Typically, one considers the moments $\< x^n \>$ which are the diagonal matrix elements of $x^n$ on an (undetermined) eigenstate $\left| E \right\>$. The moments are related by the Hamiltonian equations of motion:
\begin{equation}\label{eq:HamitonianEqQM}
    \begin{aligned}
        &\< \left[ H, \CO \right] \> = 0 \\ 
        &\< H\,\CO \> = E \< \CO \>.
    \end{aligned}
\end{equation}
It is easy to derive a recursion relation for the moments by taking $\CO$ to be $x^m$ and $x^m p$ \cite{Han:2020bkb,Berenstein:2021dyf,Bhattacharya:2021btd},
\begin{equation}\label{eq:recRelationAHO} 
    0 = m E \< x^{m-1} \> + \frac{1}{4} m(m-1)(m-2) \< x^{m-3} \> - \left\< x^m \left(\omega x + 2 g x^3 \right) \right\> 
    - 2m \left\< x^{m-1} \left(\frac{1}{2}\omega x + \frac{1}{2}g x^4 \right) \right\> \, .
\end{equation}
With this recursion an arbitrary moment $\<x^m\>$ may be expressed through combinations of $\< \mathbbm{1} \>$, $\< x \>$ and $\< x^2 \>$ up to known polynomials of $E$. The recursion relation preserves parity and parity constrains $\< x \> = 0$ so the only free parameters are $\< x^2 \>$ and $E$. Considering arbitrary $\scO = \sum_m c_m x^m,$ positivity of the norm of state $ || \CO | E \>  || \geqslant 0 $ requires $c_m c_n \< x^m x^n \> \geqslant 0$ for all real vectors $\mathbf{c}$, which is equivalent to the positive semidefiniteness of the Hankel matrix
\begin{equation}\label{eq:positivityAHO}
    M^{mn} \equiv \< x^m x^n \>, \quad \mathbf M = \left( \begin{matrix}
        1 & \< x \> & \< x^2 \> &  & \< x^K \> \\
        \< x \> & \< x^2 \> & \< x^3 \> & \cdots & \< x^{K+1} \> \\
        \< x^2 \> & \< x^3 \> & \< x^4 \> &  & \< x^{K+2} \> \\
         & \vdots & & \ddots \\
        \< x^K \> & \< x^{K+1} \> & \< x^{K+2} \> &  & \< x^{2K} \> 
    \end{matrix}\right) \succeq 0 \, ,
\end{equation}
for arbitrary truncation level $K$. 
From \ref{eq:recRelationAHO} each entry in $\mathbf{M}$ is a function only of $E$ and $\< x^2 \>$. If a particular choice of these two parameters yields $\mathbf{M} \nsucceq 0$ then it is ruled out. The parameter space shrinks with increasing $K$.
For example, taking the double-well potential $\omega^2 = -5$ and $g = 1$ the allowed parameter space is shown in Figure~\ref{fig:matrix-bootstrap}.

The matrix bootstrap application on the anharmonic oscillator is a great success, yielding rigorous bounds on the spectrum and exhibiting rapid convergence. It is natural to contemplate applications on more general dynamic observables and models. From the matrix bootstrap method all we get are constraints on the eigenvalues and diagonal matrix elements, which are useful to construct static observables. For observables that are related to decay and scattering we need the off-diagonal matrix elements. The off-diagonal matrix elements can also put stronger constraints on the existing observables. For systems having global symmetries the diagonal matrix elements are only sensitive to the Casimir elements, and all operators carrying charge will manifestly vanish. The off-diagonal matrix elements are sensitive to the global symmetry algebra. In \cite{Berenstein:2021dyf} it was reported that the above setup does not bound the ground state energy of the hydrogen atom, which may be addressed by implementing the off-diagonal matrix elements\footnote{The same authors addressed the hydrogen atom ground state energy unboundedness problem in a following-up work \cite{Berenstein:2022ygg}, using the fact that the radial wave functions of hydrogen atom states satisfy a stronger moment problem because the functions are only defined on a semi-infinite space. We conjecture that implementing the off-diagonal matrix elements may be an alternative solution to this problem. }.

It is also tempting to use the method to study models with infinite degrees of freedom, such as the lattice theory, see \cite{han2020quantum,Lawrence:2021msm} for pioneering works. In the thermodynamic limit the energy eigenvalues $E$ are dominated by IR-divergent vacuum energy and the energy density for all low energy states are the same. This means it is not easy to find a bootstrap bound on a specific low energy state\footnote{The difficulty is also technical. In the equations of motion (\ref{eq:HamitonianEqQM}) in the matrix bootstrap setup, $\< H\,\CO \> = E \< \CO \>$ is difficult to use because $E$ diverges in the thermodynamic limit and the product $H\CO$ is non-local. The other equation of motion $\< \left[ H, \CO \right] \> = 0$ does not refer to any energy eigenvalues.}. This is in contrast with the similar lattice bootstrap method  \cite{Anderson:2016rcw,Kazakov:2022xuh,Cho:2022lcj} in Euclidean spacetime and using the Schwinger-Dyson equations of motion. The latter provides bounds specifically on the ground state, due to the fact that infinite Euclidean space picks up the ground state. The difficulty can also be addressed by implementing off-diagonal matrix elements, as they are sensitive to the energy difference. 

Despite the importance of off-diagonal matrix elements, a naive generalization to include the off-diagonal matrix elements in the matrix bootstrap does not provide new bounds, as we show here. Generalizing the Hamiltonian equations of motion (\ref{eq:HamitonianEqQM}) to the off-diagonal case:
\begin{subequations}\label{eq:HamitonianEqQM-off}
    \begin{align}
        \label{eq:HamitonianEqQM-off-commutator}
        &\< E_1 | \left[ H, \CO \right] | E_2 \> = (E_1 - E_2) \< E_1 | \CO | E_2 \> \\ 
        \label{eq:HamitonianEqQM-off-nolocal}
        &\< E_1 | H\,\CO | E_2 \> = E_1 \< E_1 | \CO | E_2 \> \, .
    \end{align}
\end{subequations}
Similar to the diagonal case (\ref{eq:recRelationAHO}) one can derive a recursion relation that relates all even moments to $\<E_1| x^2 |E_2\>$ and all odd moments to $\<E_1| x |E_2\>$. The identity moment $\<E_1| \mathbbm{1} |E_2\>$ vanishes. $\<E_1| x^2 |E_2\>$ is non-vanishing only when the two external states have the same parity, and $\<E_1| x |E_2\>$ is non-vanishing only when they have opposite parity.
Therefore we can try to implement the positivity condition $ || \alpha \CO | E_1 \> + \beta \CO' | E_2 \>  || \geqslant 0 $ generalizing (\ref{eq:positivityAHO}) by:
\begin{equation}\label{eq:positivityAHO-off}
    M^{mn}_{IJ} \equiv \< E_I | x^m x^n | E_J \>, \quad \mathbf M = \left( \begin{matrix}
        \mathbf M_{11} & \mathbf M_{12}  \\
        \mathbf M_{21} & \mathbf M_{22}
    \end{matrix}\right) \succeq 0 \, .
\end{equation}
However, any $(E_1,E_2,\<E_1| x^2 |E_1\>,\<E_2| x^2 |E_2\>)$ that already satisfy (\ref{eq:recRelationAHO}) and (\ref{eq:positivityAHO}) will be trivially consistent with (\ref{eq:HamitonianEqQM-off}) and (\ref{eq:positivityAHO-off}) assigning $\< E_1 | \CO | E_2 \> = 0$. 
This obstruction motivated our search for constraints that are sensitive to the physics of transition amplitudes. The spectral bootstrap is our approach.

In the course of preparing this manuscript, \cite{Lin:2023owt} demonstrated another route toward the off-diagonal data via a Cauchy-Schwartz inequality.

\subsection{Spectral bootstrap: single matrix element bootstrap}

\begin{figure}
    \centering
    \includegraphics{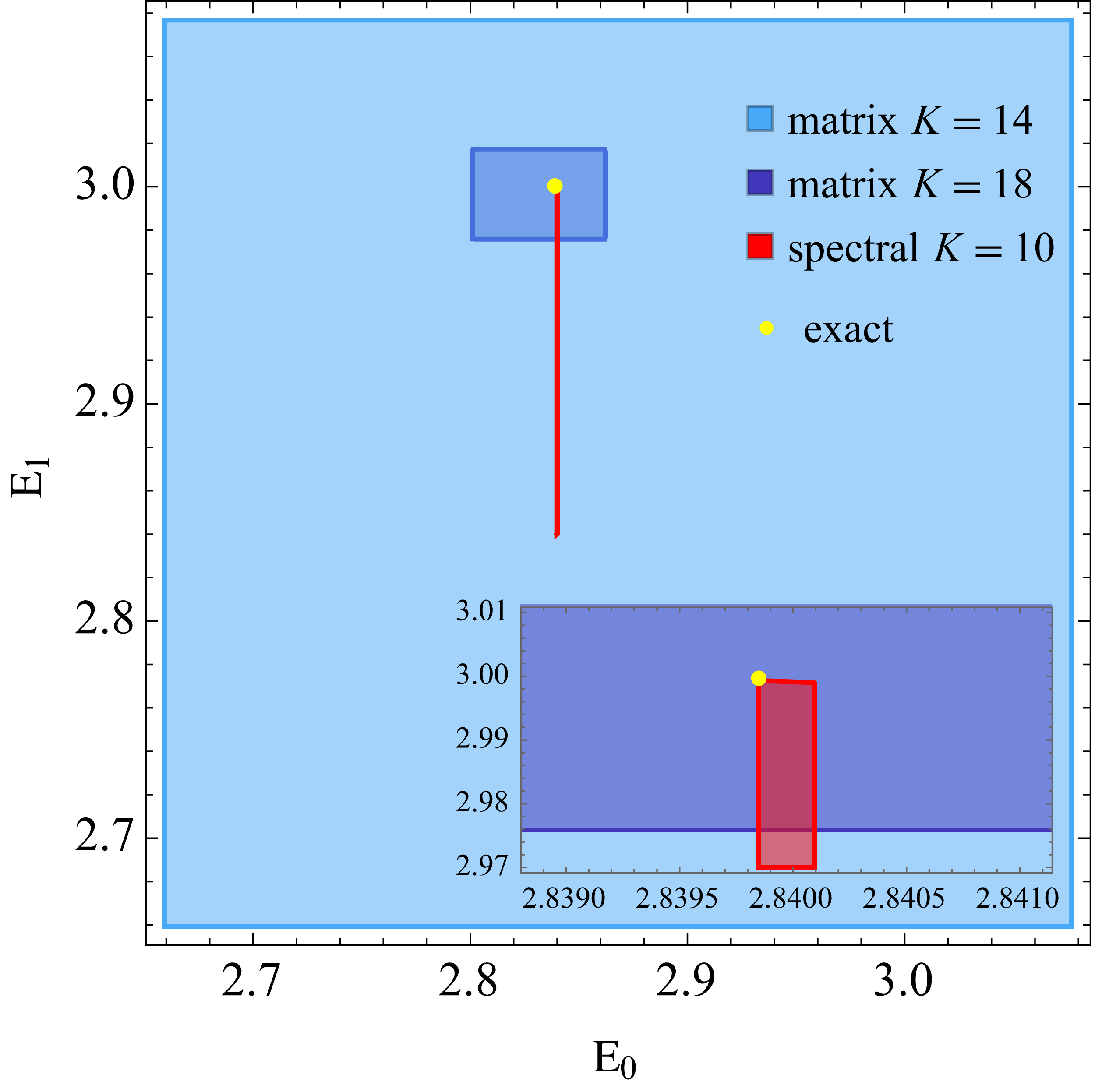}
    \caption{Comparing the bounds on the lowest two eigenvalues $E_1$ and $E_0$ from the single matrix element spectral bootstrap and the matrix bootstrap. The shaded regions are allowed by the corresponding bootstrap setup. The allowed region of on $E_0$ is very narrow, and the inset plot zooms in to the vicinity of the exact answer. The single matrix element spectral bootstrap furnishes only an upper bound on $E_1$, which is saturated by the exact value.}
    \label{fig:single-AHO}
\end{figure}

In this subsection we introduce the spectral bootstrap in its simplest setup. As is summarized in Table~\ref{tab:analogy}, the spectral bootstrap shares many similarities with the conformal bootstrap, and we will use the analogy to guide our introduction. The key structure of the conformal bootstrap is the crossing equations. In spectral bootstrap we consider the ground state diagonal matrix element $\left\< 0 \left| x^i x^j \right| 0 \right\>$, insert a complete set of states between $x^i$ and $x^j$, and obtain our crossing equations
\begin{equation}\label{eq:crossing-equation-AHO}
    \left\< 0 \left| x^{i+j} \right| 0 \right\> =  \left\< 0 \left| x^i \right| 0 \right\> \left\< 0 \left| x^j \right| 0 \right\> + \sum_{k\neq 0} \left\< 0 \left| x^i \right| k \right\> \left\< k \left| x^j \right| 0 \right\> \, .
\end{equation}
Schematically, the l.h.s. describes the ``t-channel'' where one operator fuses with the other operator and acts on the external states, while the r.h.s describes the ``s-channel'' where the operator fuses with the external state and creates the whole spectrum through the off-diagonal matrix elements. The latter is reminiscent of the operator product expansion (OPE) in the conformal bootstrap, and state-operator-state matrix elements are the ``OPE coefficients''. Here we introduce a more economical notation that emphasizes the similarity with the CFT bootstrap:
\begin{equation}
c_{k\ell,\scO} \equiv \< k \left| \scO \right| \ell \> \, .
\end{equation}
Just like the OPE in a CFT can be classified into primary operators and their descendants whose OPE coefficients are completely fixed up to the primary OPE coefficients by conformal symmetry, in the spectral bootstrap case the equations of motion (\ref{eq:HamitonianEqQM}) and (\ref{eq:HamitonianEqQM-off}) fix many matrix elements. It turns out that the only ``primary operators'' are $x$ and $x^2$, and everybody else is a descendant whose matrix elements are fixed up to the primary matrix elements
\begin{equation}\label{eq:descendants-AHO}
\< k \left| x^n \right| \ell \> = \delta_{k\ell} g_{n}^{(\mathbbm{1})}(E_k) + c_{kl,x} g_{n}^{(x)}(E_k,E_\ell) + c_{kl,x^2} g_{n}^{(x^2)}(E_k,E_\ell) \, ,
\end{equation}
where $g_{n}^{(\CO_{\rm prim})}(E_k)$ denotes some known polynomials of $E_k$ and $E_\ell$. If we substitute (\ref{eq:descendants-AHO}) in the s-channel spectral expansion in the crossing equations (\ref{eq:crossing-equation-AHO}), we further put the crossing equations into the ``conformal block decomposition'' form 
\begin{align}
\label{eq:s-channel-decompose-AHO-no-symmetry-0}
\left\< 0 \left| x^i \right| 0 \right\> \left\< 0 \left| x^j \right| 0 \right\> &= 
    (\begin{matrix} 1 & c_{00,x^2} \end{matrix})
\big( \vec{\CS}_0 \big)_{ij}
\left(\begin{matrix} 1 \\ c_{00,x^2} \end{matrix}\right)
\\
\label{eq:s-channel-decompose-AHO-no-symmetry-k}
\quad \sum_{k\neq 0} \left\< 0 \left| x^i \right| k \right\> \left\< k \left| x^j \right| 0 \right\> &= 
\sum_k 
(\begin{matrix} c_{0k,x} & c_{0k,x^2} \end{matrix})
\big( \vec{\CS}_k \big)_{ij}
\left(\begin{matrix} c_{0k,x} \\ c_{0k,x^2} \end{matrix}\right)
\end{align}
where the vectors $\vec{\CS}_0$ and $\vec{\CS}_k$ contain the ``conformal blocks'', 
\begin{align}
    \big( \vec{\CS}_0 \big)_{ij} &\equiv \left(\begin{matrix}
        g_{i}^{(\mathbbm{1})}(E_0) g_{j}^{(\mathbbm{1})}(E_0) & g_{i}^{(\mathbbm{1})}(E_0) g_{j}^{(x^2)}(E_0) \\
        g_{i}^{(x^2)}(E_0) g_{j}^{(\mathbbm{1})}(E_0) & g_{i}^{(x^2)}(E_0) g_{j}^{(x^2)}(E_0)
    \end{matrix}\right) 
\\
    \big( \vec{\CS}_k \big)_{ij} &\equiv \left(\begin{matrix}
        g_{i}^{(x)}(E_0,E_k) g_{j}^{(x)}(E_k,E_0) & g_{i}^{(x)}(E_0,E_k) g_{j}^{(x^2)}(E_k,E_0) \\
        g_{i}^{(x^2)}(E_0,E_k) g_{j}^{(x)}(E_k,E_0) & g_{i}^{(x^2)}(E_0,E_k) g_{j}^{(x^2)}(E_k,E_0)
    \end{matrix}\right)  \, .
\end{align}
The block expansion can be further decomposed into the irreducible representations of the global symmetries. In the anharmonic case the symmetry is the parity-$\mathbb{Z}_2$. For parity even states $c_{0k,x}$ vanish and for parity odd states $c_{0k,x^2}$ vanish, so we can further decompose the ($\ref{eq:s-channel-decompose-AHO-no-symmetry-k}$) into the parity subsectors 
\begin{equation}\label{eq:spectral-s-channel-AHO}
\begin{aligned}
    \sum_k &
        (\begin{matrix} c_{0k,x} & c_{0k,x^2} \end{matrix})
        \vec{\CS}_k 
        \left(\begin{matrix} c_{0k,x} \\ c_{0k,x^2} \end{matrix}\right)
    = \sum_{k_-} c_{0k_-,x}^2 \vec{\CS}_{k_-} + \sum_{k_+} c_{0k_+,x^2}^2 \vec{\CS}_{k_+} \\ 
    & \big( \vec{\CS}_{k_-} \big)_{ij} \equiv g_{i}^{(x)}(E_0,E_k) g_{j}^{(x)}(E_k,E_0) \\ 
    & \big( \vec{\CS}_{k_+} \big)_{ij} \equiv g_{i}^{(x^2)}(E_0,E_k) g_{j}^{(x^2)}(E_k,E_0)
\end{aligned}
\end{equation}
where $k_-$ and $k_+$ are indices of parity odd and even states, respectively. Now we switch to the t-channel in the crossing equation (\ref{eq:crossing-equation-AHO}) and apply the same descendant rules (\ref{eq:descendants-AHO}) as in the s-channel
\begin{equation}\label{eq:spectral-t-channel-AHO}
    \left\< 0 \left| x^{i+j} \right| 0 \right\> = 
    \big(\begin{matrix} 1 & c_{00,x^2} \end{matrix}\big)
    \left(\begin{matrix}
        g_{i+j}^{(\mathbbm{1})}(E_0) 
        & \frac{1}{2} g_{i+j}^{(x^2)}(E_0) 
        \\
        \frac{1}{2} g_{i+j}^{(x^2)}(E_0) 
        & 0
    \end{matrix}\right)
    \left(\begin{matrix} 1 \\ c_{00,x^2} \end{matrix}\right)
    \equiv 
    \big(\begin{matrix} 1 & c_{00,x^2} \end{matrix}\big)
    \big( \vec{\CT}_0 \big)_{ij}
    \left(\begin{matrix} 1 \\ c_{00,x^2} \end{matrix}\right) \, .
\end{equation}
Notice that even though the t-channel is linear in the matrix elements, one can still embed it in the bilinear form because the $c$-vector contains a constant.
Equating (\ref{eq:spectral-t-channel-AHO}) and (\ref{eq:spectral-s-channel-AHO}), we obtain the crossing equations
\begin{equation}\label{eq:spectral-crossing-AHO}
    0 = \big(\begin{matrix} 1 & c_{00,x^2} \end{matrix}\big) \big(\vec{\CS}_0 - \vec{\CT}_0\big)
    \left(\begin{matrix} 1 \\ c_{00,x^2} \end{matrix}\right) + 
    \sum_{k_-} c_{0k_-,x}^2 \vec{\CS}_{k_-} + \sum_{k_+} c_{0k_+,x^2}^2 \vec{\CS}_{k_+} \, .
\end{equation}
The crossing equations (\ref{eq:spectral-crossing-AHO}) work in exactly the same way as those in the conformal bootstrap. Because the coefficients $c_{0k,\CO}$ are real and show up in a quadratic form, we can use semidefinite programming to test spectral assumptions. It is natural to assume the first excitation state is at $E_1$ which is above the ground state energy $E_0$, and all other energies are above $E_1$. This defines the feasibility problem\footnote{In principle one can require $\big(\vec{\CS}_0 - \vec{\CT}_0\big) \succeq 0$ instead and scanning $c_{00,x^2}$ is not necessary. In the anharmonic oscillator case we find that the bound is significantly stronger in (\ref{eq:bootstrap-problem-spectral-single}) with $c_{00,x^2}$ values explicitly provided.}:
\begin{equation}\label{eq:bootstrap-problem-spectral-single}
\boxed{
\begin{aligned}
&\text{If there exists $\alpha_{ij}$ such that $\forall E_{k_\pm} \geqslant E_1$}\\ 
&\sum_{i,j\leqslant K} \alpha_{i,j} \big(\begin{matrix} 1 & c_{00,x^2} \end{matrix}\big)
\big(\vec{\CS}_0 - \vec{\CT}_0\big)_{ij}
\left(\begin{matrix} 1 \\ c_{00,x^2} \end{matrix}\right) = 1 \\ 
& \sum_{i,j\leqslant K} \alpha_{i,j} \big(\vec{\CS}_{k_-}\big)_{ij} \geqslant 0 \\ 
& \sum_{i,j\leqslant K} \alpha_{i,j} \big(\vec{\CS}_{k_+}\big)_{ij} \geqslant 0 \, , \\ 
& \text{then all spectra with the prescribed ($E_0$,$E_1$) and $c_{00,x^2}$ are ruled out.}
\end{aligned}
}
\end{equation}
The numerical bootstrap bound is shown in Figure~\ref{fig:single-AHO}. In the plot we project all allowed space on the $(E_0, E_1)$ plane, and compare the bounds from the spectral bootstrap and the matrix bootstrap. The single matrix element setup (\ref{eq:bootstrap-problem-spectral-single}) can only provide upper bounds on the gap $E_1$ without additional assumptions, because any allowed $(E_0, E_1)$ means any $E_1' \leq E_1$ is trivially allowed, so the only lower bound is $E_1 \geq E_0$. Nevertheless even at modest $K$ the bound on $E_0$ in the spectral bootstrap case is much stronger than the matrix bootstrap at large $K$. The upper bound on $E_1$ is also much stronger than the matrix bootstrap as the matrix bootstrap upper bound is not saturated by the exact answer while the spectral bootstrap bound is almost saturated. It is likely that implementing an assumption on $E_2$ could enable the spectral bootstrap to produce a strong lower bound as well.

\subsection{Spectral bootstrap: mixed matrix element bootstrap}
Analogous to the mixed correlator bootstrap in the conformal bootstrap, we can promote our bootstrap problem (\ref{eq:bootstrap-problem-spectral-single}) to a mixed system by considering matrix elements $\< 0 | x^i x^j | 0 \>$, $\< 1 | x^i x^j | 1 \>$ and $\< 0 | x^i x^j | 1 \>$ assuming the first excited state $| 1 \>$ has odd parity. We write down crossing equations similar to (\ref{eq:spectral-crossing-AHO}), but for the above 3 matrix elements jointly. The mixed crossing equations contain an isolated s-channel $\vec{\CS}_{\rm c}$ involving the matrix elements between the external states $\< 0 | x^i | 0 \>$, $\< 0 | x^i | 1 \>$, $\< 1 | x^i | 1 \>$, isolated t-channel $\vec{\mathcal{T}}_{\rm c}$ and two internal channels $\vec{\CS}_-$ and $\vec{\CS}_+$ for the odd- and even-parity states above $E_1$. The mixed matrix element crossing equation can be summarized as a vector block form
\begin{equation}\label{eq:mixed-equations-AHO}
\begin{aligned}
0 &= \sum_{k_-} 
\big( \begin{matrix} c_{0k,x} & c_{1k,x^2} \end{matrix}\big) \vec{\CS}_- 
\left( \begin{matrix} c_{0k,x} \\ c_{1k,x^2} \end{matrix}\right)
+ \sum_{k_+} 
\big( \begin{matrix} c_{0k,x^2} & c_{1k,x} \end{matrix}\big) \vec{\CS}_+ 
\left( \begin{matrix} c_{0k,x^2} \\ c_{1k,x} \end{matrix}\right) \\
&\quad + \big( \begin{matrix} 1 & c_{00,x^2} & c_{11,x^2} & c_{01,x} \end{matrix}\big)
\big( \vec{\CS}_{\rm c} - \vec{\mathcal{T}}_{\rm c} \big)
\big( \begin{matrix} 1 & c_{00,x^2} & c_{11,x^2} & c_{01,x} \end{matrix}\big)^{T} \, .
\end{aligned}
\end{equation}
The details of the block vectors $\vec{\CS}_\pm$, \, $\vec{\CS}_c$ and $\vec{\CT}_c$ are explained in appendix~\ref{sec:mix-details-AHO}.

\begin{figure}[htbp]
    \centering
    \includegraphics[width=0.6\linewidth]{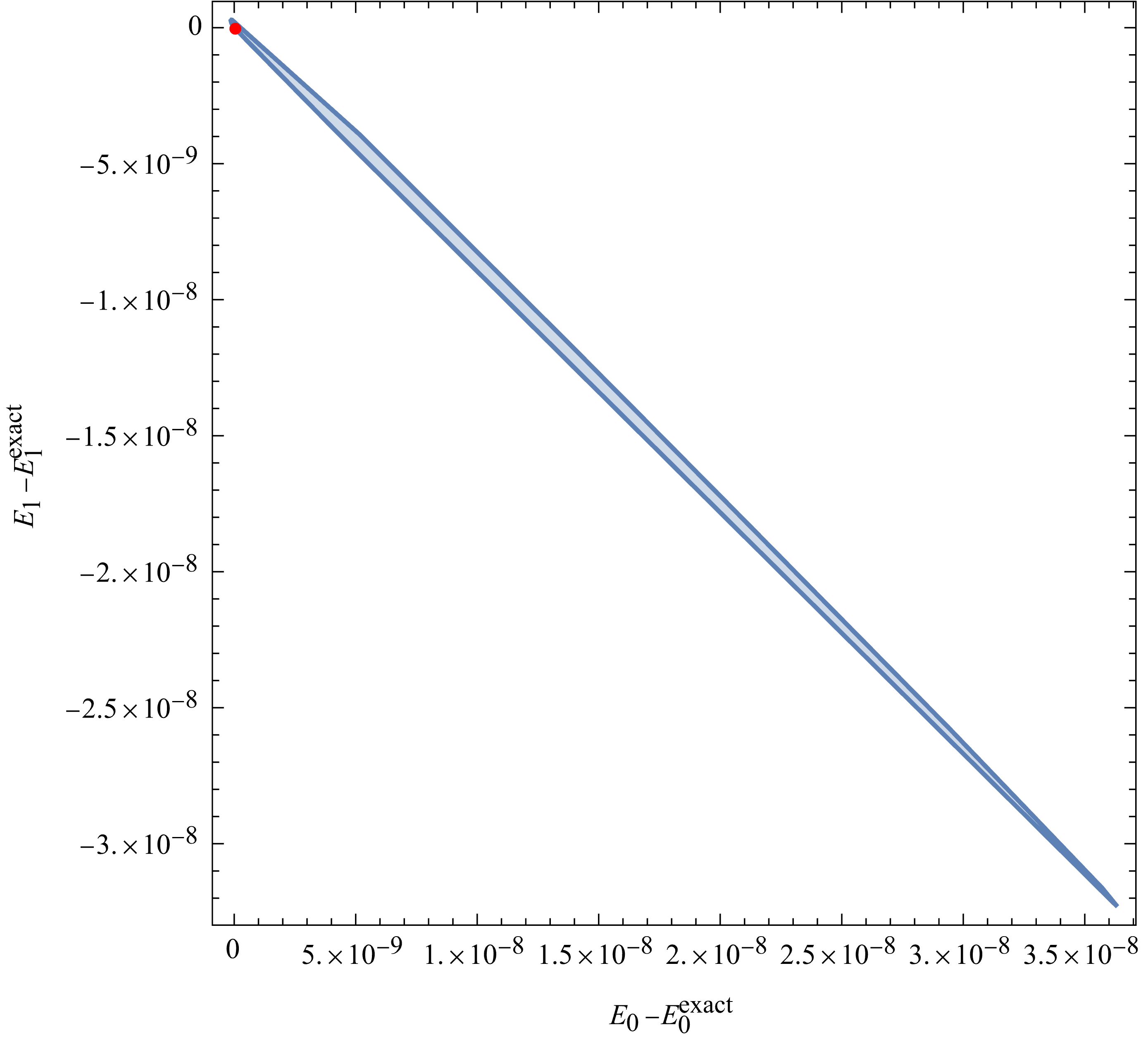}
    \caption{Bound on the first two eigenvalues $(E_0, E_1)$ from the mixed matrix element bootstrap equation (\ref{eq:mixed-equations-AHO}) at $K=8$. The red marker represents the exact $(E_0, E_1)$ values. The blue shaded region is the allowed parameter space of $(E_0, E_1)$. The region is a projection from the 5-dimensional parameter space $(E_0, E_1, c_{00,x^2} , c_{11,x^2} , c_{01,x})$. Since the 5-dimensional sampling can be lossy, the true allowed region may be larger than the island shown in the plot. 
    }
    \label{fig:island-mixed}
\end{figure}
We show the bound on the first two eigenvalues $(E_0, E_1)$ in Figure~\ref{fig:island-mixed}. The allowed parameter space is a thin needle-like island whose one tip is saturated by the exact solution with an error $\lesssim 10^{-10}$. The island is also many orders of magnitude smaller than the single matrix element bound in Figure~\ref{fig:single-AHO}, despite the fact that the mixed matrix element bootstrap is at smaller $K$. The island is a projection of the 5-dimensional parameter space $(E_0, E_1, c_{00,x^2} , c_{11,x^2} , c_{01,x})$, computed by sampling about 500 angular directions on $S^5$. For each direction we find the boundary using a navigator function\cite{Reehorst:2021ykw} and line search between an allowed point and a disallowed point. Such an algorithm is not ideal for sampling a 5-dimensional space. In the future we can obtain a more reliable island using the OPE scan method\cite{Chester:2019ifh}. Nevertheless the current numerical result is enough to support our point that the mixed matrix element bootstrap imposes much stronger bounds on the anharmonic oscillator spectrum.
\begin{figure}[htbp]
    \centering
    \includegraphics[width=0.45\linewidth]{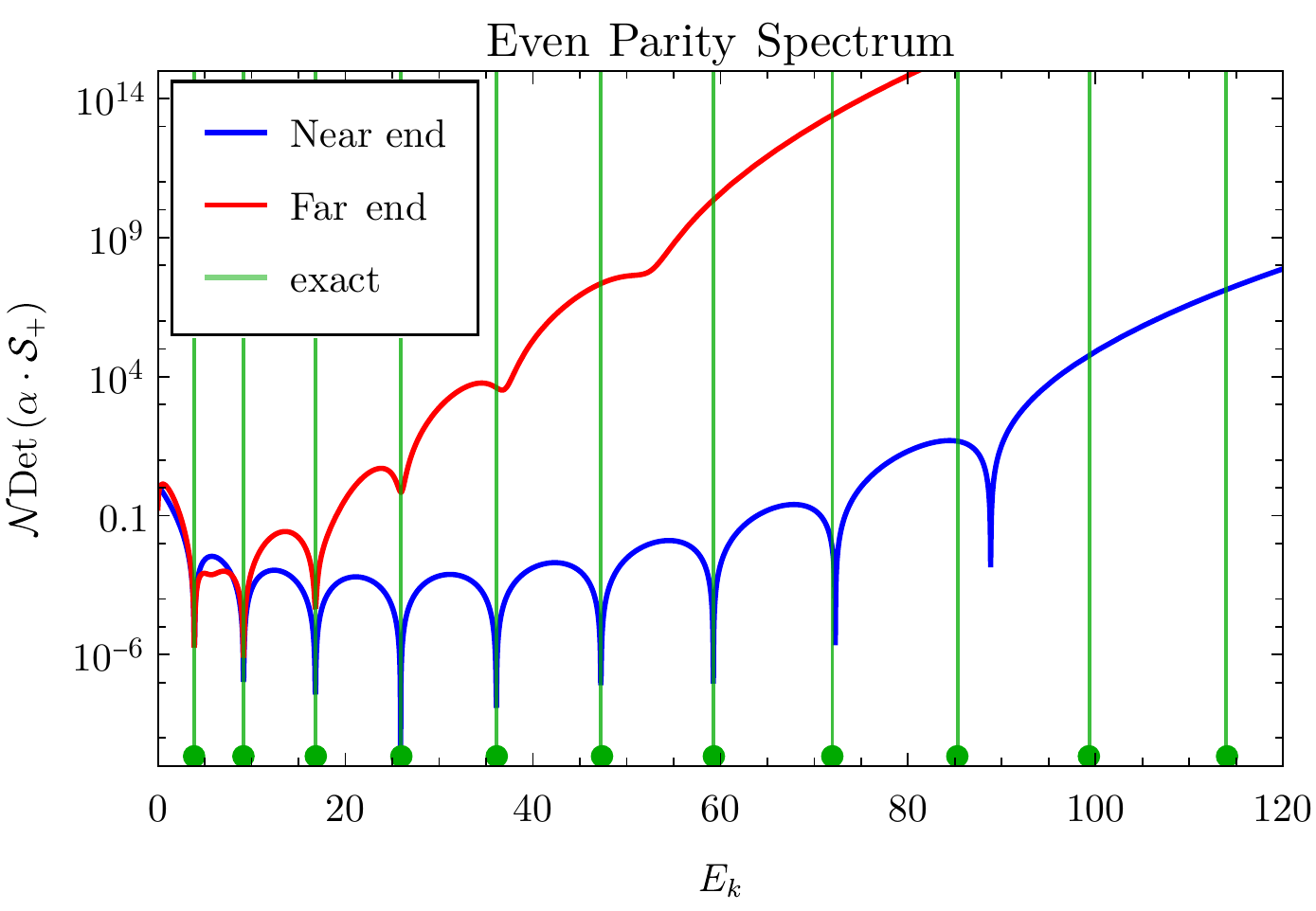}
    \includegraphics[width=0.45\linewidth]{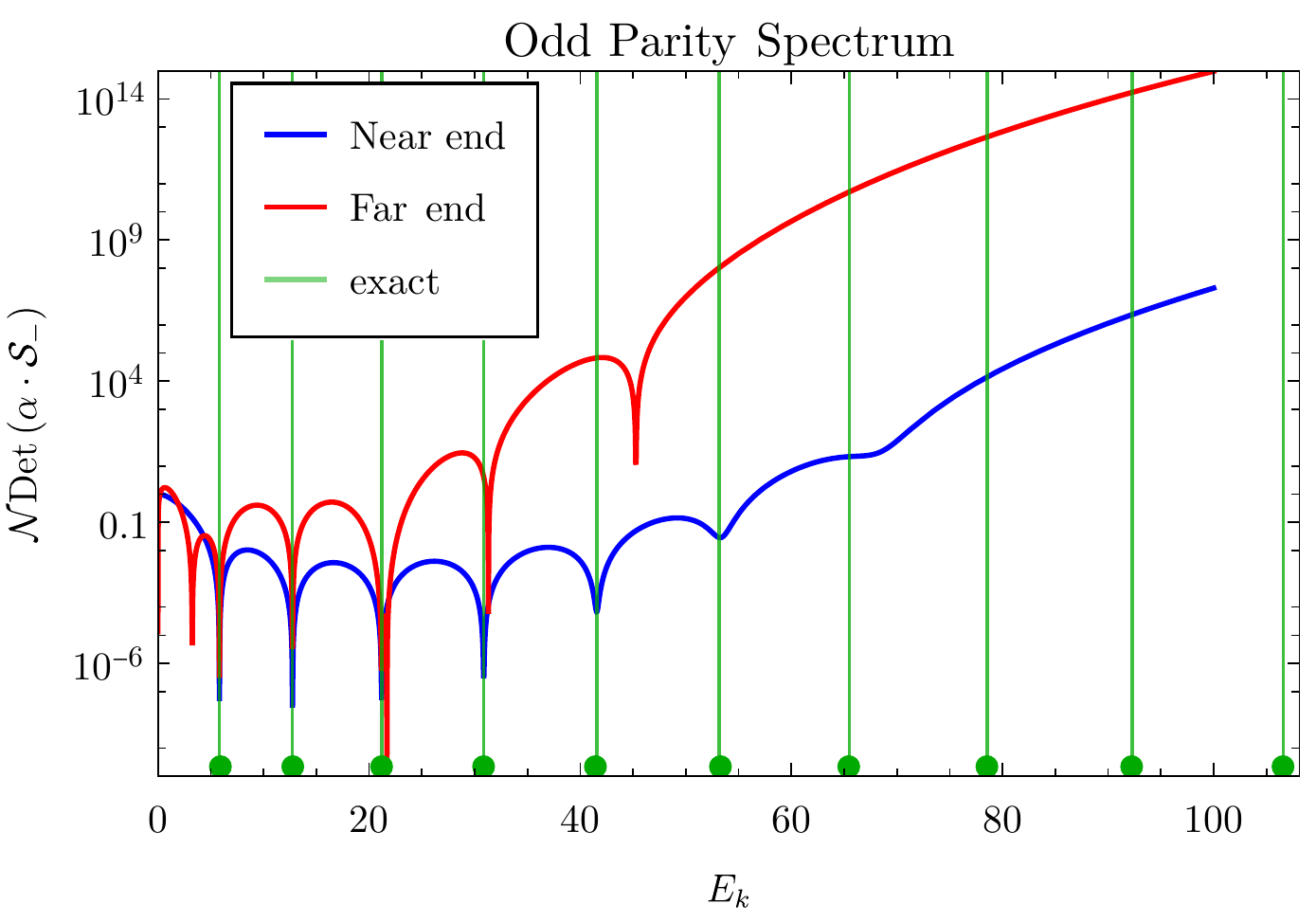}
    \caption{Extremal functional of the mixed matrix element bootstrap extremal bound. The blue and red colors correspond to two tips of the island, that is one ``near end'' at the tip of the island near the exact solution, and another ``far end'' at the other tip of the island. Green vertical lines represent the exact eigenvalues of the excited states. The positions of double zeros in the extremal functional is an approximation of the excited states. The absolute magnitude of the extremal functional is irrelevant, and we multiply each functional by a normilization constant $\mathcal{N}$ in order to show them in the same plot.
    }
    \label{fig:efm}
\end{figure}
When the bounds on the external parameters $(E_0, E_1)$ are nearly saturated the bootstrap equations also strongly constrain higher excitations. We can extract an approximate spectrum of the excited states using the extremal functional method\cite{El-Showk:2012vjm,Simmons-Duffin:2016wlq}. Schematically, when we go to the parameter point that is marginally ruled out, the crossing equation (\ref{eq:mixed-equations-AHO}) and the non-negativity of the functional $\alpha \cdot \CS$ must be simultaneously satisfied, and a consistent functional satisfying both will need to have double roots at the energy eigenvalues of the excited states. We show the extremal functional plots of the two tips of the island in Figure~\ref{fig:efm}. In the plots we find that in both ``near end'' and ``far end'' cases the double zeros line up accurately with the exact spectrum. The near end functional has more double zeros which agree perfectly with the exact spectrum. 

The results in this section show that the spectral bootstrap imposes highly non-trivial constraints on the anharmonic oscillator model. Next we are ready to generalize this setup to study infinite lattice.

\section{Infinite lattice example: Transverse Field Ising Model}\label{sec:tfim}
The one-dimensional transverse-field Ising model (TFIM) is a quantum spin chain with the hamiltonian
\begin{equation}\label{eq:TFIM-hamiltoian}
    H_{\mathrm{TFIM}} = -\sum_{i=1}^N\sigma_i^z\sigma_{i+1}^z + h \sum_{i=1}^N \sigma_i^x
\end{equation}
where $\sigma_i^{x,y,z}$ denote the Pauli matrices associated to site $i$ of a (possibly infinite) one-dimensional lattice. 
This model has a quantum phase transition with a critical point $h=1$ separating a $Z_2$ symmetry-broken phase and a paramagnetic symmetry-unbroken phase. Its physics are readily accessible through exact diagonalization, density renormalization group, or quantum monte carlo. It can be diagonalized analytically via a Jordan-Wigner transformation that maps it into a theory of free fermions, up to Bogoliubov transformation. As a well-understood model with infinitely many degrees of freedom amenable solution both numerically and analytically, the TFIM provides a sound basis for an assessment of the spectral bootstrap method in the many-body regime.

We will briefly sketch the procedure for solving the TFIM analytically, see \cite{sachdev_2011} for a complete discussion. The first step is to rewrite the spin operators in terms of fermion operators, which is called Jordan-Wigner transformation:
\begin{align}\label{eq:JordanWigner}
    \sigma_i^+ &= \prod_{j<i}\left(1-2c_j^\dagger c_j\right)c_i\\
    \sigma_i^- &= \prod_{j<i}\left(1-2c_j^\dagger c_j\right)c_i^\dagger
\end{align}
where $\sigma_i^\pm = (\sigma_i^x \pm \mathrm{i}\sigma_i^y)/2.$ The second step is to employ the fourier representation of the fermions
\begin{equation}
    c_k = \frac{1}{\sqrt{N}}\sum_j c_j \mathrm{e}^{-\mathrm{i} k x_j}.
\end{equation}
Finally a Bogoliubov transformation
\begin{equation}
    \gamma_k = u_k c_k - \mathrm{i}v_k c^\dagger_{-k}
\end{equation}
maps the problem into a theory of free fermions. Having omitted the details, we only want to emphasize that the basis $\gamma_k$ which diagonalizes the model is on account of \ref{eq:JordanWigner} highly nonlocal in the spin variables $\sigma_i$. It may prove fruitful to study other theories using the spectral bootstrap in the fermion basis, but since we would like to avoid trivializing the computational problem for this integrable model we will use only the spin basis in the present study. From the analytical approach we will quote the known spectrum and magnetization.

In the free fermion picture the hamiltonian for the infinite system is
\begin{equation}H_{\mathrm{TFIM}} = \sum_k \epsilon_k (\gamma_k^\dagger \gamma_k + 1/2)\end{equation}
with dispersion
\begin{equation}\epsilon_k = 2 \sqrt{1+h^2 - 2 h \cos{k}}.\end{equation}
For our purposes we need only the energy difference between the lowest two eigenstates, given by
\begin{equation}
    E_1-E_0 = \begin{cases}
        0, & h < 1 \\
        2|h-1| & h > 1
    \end{cases}.
\end{equation}
Placing bounds on this quantity using the spectral bootstrap is the main object of this section. We will also compute it using exact diagonalization of the spin hamiltonian at finite $N$ to have another numerical method to which we can compare our own.

Computation of the magnetization is more involved than that of the energy spectrum, however it has been obtained as a simple integral formula \cite{PhysRev.127.1508}
\begin{equation}
    \< \sigma^x \> =
        \frac{1}{\pi}\lim_{\beta \rightarrow \infty}\int_0^\pi \frac{(\cos (\omega )-h) \tanh \left(\sqrt{\beta  \left(h^2-2 h \cos (\omega )+1\right)}\right)}{  \sqrt{h^2-2 h \cos (\omega )+1}} \, \mathrm{d} \omega .
\end{equation}

\subsection{Setup for infinite lattice}
In this subsection we introduce the spectral bootstrap setup for TFIM. Schematically, we begin with the crossing equations $\< 0 | \CO \CO' | 0 \> = \sum_k \< 0 | \CO |k\> \< k |\CO' | 0 \>$, then use the Hamiltonian equations of motion and global symmetries to reduce the matrix elements $\< 0 | \CO |k\>$ into a small primary set, and finally use semidefinite programming to test spectral assumptions. As is discussed in section \ref{sec:matrix-bootstrap}, the infinite lattice has subtleties and we will need to adjust our bootstrap setup. We do not use the equations of motion of the product type (\ref{eq:HamitonianEqQM-off-nolocal}) because $E$ is IR-divergent. The rest of the equations of motion are of the commutator type (\ref{eq:HamitonianEqQM-off-commutator}). Unlike the quantum mechanical case, matrix elements $\< 0 | \CO |k\>$ do not reduce to a finite primary set merely using the commutator equations so we must truncate the number of primaries. Our strategy is to take the primary operator set $\mathcal{A}$ to be a small set of local ``string operators'' $\CO \in \mathcal{A} \subset \{ \sigma_1^x, \sigma_1^z, \sigma_1^x \sigma_2^x, \sigma_1^x \sigma_2^z , \cdots \}$ and generate the full set of operators by recursively commuting with the Hamiltonian
\begin{subequations}\label{eq:time-derivative-and-matrix-element}
    \begin{equation}\label{eq:time-derivative}
        \pd \CO \equiv i\left[ H, \CO \right], \quad \pd^{n+1} \CO \equiv i\left[ H, \pd^n\CO \right].
    \end{equation}
    The generated operators are manifestly descendant operators that do not have independent matrix elements:
    \begin{equation}
        \< \ell | \pd^{n} \CO | k \> = i^n (E_\ell - E_k)^n \< \ell | \CO | k \> \, .
    \end{equation}
\end{subequations}
Similar to the Hamiltonian equations of motion, translational invariance offers additional equations of motion. We define a unitary operator $U$ that permutes all sites by 1 lattice unit, and all eigenstates acquire a definitive phase under permutation
\begin{equation}
    U \vec\sigma_i = \vec\sigma_{i-1} U, \quad U | E,p \> = e^{i p} | E,p \> \, .
\end{equation}
This allows us to generate an extended set of descendant operators
\begin{subequations}\label{eq:spacial-derivative-and-matrix-element}
\begin{align}\label{eq:spacial-derivative}
    \nabla^{n} \CO &\equiv U^n\CO U^{-n} \\
    \< \ell | \nabla^{n} \CO | k \> &= e^{in(p_\ell - p_k)} \< \ell | \CO | k \> \,.
\end{align}
\end{subequations}
The $\log$ of the phase is identified as the momentum of the states. For a lattice the momentum is only defined on the first Brillouin zone $-\pi \leq p \leq \pi$. On an infinite lattice, $p$ is continuous.
A primary operator and its descendants form a multiplet
\begin{equation}\label{eq:fusion}
    \vec{\mathcal{D}}\CO := \{ \CO, \pd \CO, \nabla \CO, \pd^2 \CO, \pd \nabla \CO , \cdots , \nabla^\Lambda \CO \}
\end{equation}
and we truncate each multiplet to contain all descendants with at most $\Lambda$ total number of derivatives. 
The crossing equations can be generated from the fusion between these multiplets
\begin{equation}\label{eq:multiplets}
    \vec{\mathcal{D}}\CO \times \vec{\mathcal{D}}\CO' \rightarrow \vec{\mathcal{D}}\CO'' + \vec{\mathcal{D}}\CO''' + \cdots \, .
\end{equation}
In the lattice theory we have an infinite number of primaries, and we truncate the primaries and their multiplets. This means when we fuse $\vec{\mathcal{D}}\CO$ with $\vec{\mathcal{D}}\CO'$, we may get operators that do not belong to any truncated multiplets that we consider. Such operators do not provide useful bounds on the spectrum, so we take a subset of crossing equations that eliminates those operators. The most general form of our crossing equations is the following
\begin{equation}\label{eq:crossing-equations-general-TFIM}
    \begin{aligned}
        &\sum_{i'j'} T_{I,i'j'} \left\< E_{m},p_m | \mathcal{D}_{i'}\CO_{j'} | E_m, p_m \right\> \\
        =&\sum_{ijkl} S_{I,ijkl} \sum_n \left\< E_{m},p_m | \mathcal{D}_i\CO_j | E_n, p_n \right\> \left\< E_n, p_n | \mathcal{D}_k\CO_\ell | E_m, p_m \right\>
    \end{aligned}
\end{equation}
where $I$ is the index of the crossing equations, and $T_{I,i'j'}$ and $S_{I,ijkl}$ are coefficients that are determined by fusion (\ref{eq:fusion}) and linear combinations that eliminate unbounded operators. We currently search for the crossing equations by brute force. The advantage of the setup in this subsection is that the number of undetermined matrix elements are minimized, and in the semidefinite programming step we will have polynomial matrices whose dimensions are minimal.

\subsection{Reduction by symmetry}
The Hamiltonian (\ref{eq:TFIM-hamiltoian}) enjoys spin-$\mathbb{Z}_2$ symmetry, parity symmetry and time reversal symmetry. There is no continuous global symmetry. These discrete symmetries can be used to further classify the primary operators and their matrix elements. 

The spin-$\mathbb{Z}_2$ symmetry is generated by 
\begin{equation}
    S = \prod_i \sigma_i^x \, .
\end{equation}
We can classify all operators according to their spin, and they transform under $S$ as $S\CO S^{-1} = s_\CO \CO$ where $s_\CO = \pm 1$. Since $S$ also commutes with translation $U$ and reflection $P$, we can take each eigenstate to have definite spin. The matrix elements have the spin-selection rule 
\begin{equation}\label{eq:spin-selection-rule}
    \< s | \CO | s' \> = \< s | S^{-1} S \CO S^{-1} S | s' \> = ss's_\CO \< s | \CO | s' \>
\end{equation}
indicating that the matrix element vanishes when the product $s s' s_\scO$ is $-1$.

We define the reflection operation as reflecting all spins about site $1$:
\begin{equation}
    P \vec{\sigma}_i P^{-1} = \vec{\sigma}_{2-i} \, .
\end{equation}
We symmetrize and anti-symmetrize the operator 
to obtain operators with definite parity 
\begin{equation}
    \begin{aligned}
        \CO^{(\pm)} &\equiv \frac{1}{2} \left( \CO \pm P \CO P^{-1} \right) 
    \end{aligned}
\end{equation}
There is a subtlety in the parity selection rule. We would like to take our eigenstates to have definite momentum in order to take advantage of translational invariance, but the direction of momentum is not invariant under reflection. So we instead label each state with $| E, |p|, r, s \>$ keeping the energy $E$, the absolute value of momentum $|p|$, the parity $r$ and the spin $s$ to be definite. 
With this assignment, matrix elements that do not have overall even parity will vanish.
For matrix elements of descendant operators between states carrying nonzero momenta, they are fixed as a mixture of even- and odd- parity primary matrix elements according to the following rule: 
\begin{equation}\label{eq:parity-selection-rule}
\begin{aligned}
    & \left(
        \begin{array}{c} 
            \< p, + | (\nabla^n \CO)^{(+)} | p', + \> \\ 
            \< p, - | (\nabla^n \CO)^{(+)} | p', - \> \\
            \< p, + | (\nabla^n \CO)^{(-)} | p', - \> \\
            \< p, - | (\nabla^n \CO)^{(-)} | p', + \> 
        \end{array}
    \right) = Y(n,p,p') \left(
        \begin{array}{c} 
            \< p, + | \CO^{(+)} | p', + \> \\ 
            \< p, - | \CO^{(+)} | p', - \> \\
            \< p, + | \CO^{(-)} | p', - \> \\
            \< p, - | \CO^{(-)} | p', + \> 
        \end{array}
    \right) \\
    & Y(n,p,p') = \begin{scriptsize}
        \left(
            \begin{array}{cccc}
             \cos (n p) \cos \left(n p'\right) & \sin (n p) \sin \left(n p'\right) & i \cos (n p)
               \sin \left(n p'\right) & -i \sin (n p) \cos \left(n p'\right) \\
             \sin (n p) \sin \left(n p'\right) & \cos (n p) \cos \left(n p'\right) & -i \sin (n p)
               \cos \left(n p'\right) & i \cos (n p) \sin \left(n p'\right) \\
             i \cos (n p) \sin \left(n p'\right) & -i \sin (n p) \cos \left(n p'\right) & \cos (n p)
               \cos \left(n p'\right) & \sin (n p) \sin \left(n p'\right) \\
             -i \sin (n p) \cos \left(n p'\right) & i \cos (n p) \sin \left(n p'\right) & \sin (n p)
               \sin \left(n p'\right) & \cos (n p) \cos \left(n p'\right) \\
            \end{array}
        \right)
    \end{scriptsize}
\end{aligned} \, ,
\end{equation}
where $E$ and $s$ indices of the states are suppressed.

Finally the phases of the matrix elements are fixed by time-reversal symmetry. The action of the time reversal operator $T$ is simply complex conjugation\footnote{We thank Meng Cheng for explaining the time-reversal symmetry in the TFIM case in \url{https://physics.stackexchange.com/questions/228821/time-reversal-symmetry-of-transverse-field-ising-model} }, so 
\begin{equation}
    T\sigma^{x,z} T^{-1} = \sigma^{x,z}, \quad T\sigma^{y} T^{-1} = -\sigma^{y} \, .
\end{equation}
Complex conjugation acts trivially on the translation operator $U$, reflection operator $P$ and the spin-$\mathbb{Z}_2$ generator, but reflects the momentum as
\begin{equation}
    U T \left| E, p, s \right \> = T U \left| E, p, s \right \> = T e^{ip} \left| E, p, s \right \> = e^{-ip} T \left| E, p, s \right \> \,.
\end{equation}
Since $T$ is merely complex conjugation it satisfies $T^2 = 1$ and there is no Kramers degeneracy. Without loss of generality we can choose an eigen-basis that diagonalizes $PT$ such that $PT | E, |p|, r, s \> = | E, |p|, r, s \>$. One can prove that each matrix element is either real or imaginary by
\begin{equation}\label{eq:phase-selection-rule}
    \< i | \CO | f \> = \< f | (PT)\CO(PT)^{-1} | i \> = 
    \pm \< i | \CO | f \>^{*}
\end{equation}
where the $\pm$ sign are determined by the operator's parity and number of $i$'s and $\sigma^y$'s. In particular, $\pd$ as defined in (\ref{eq:time-derivative})  flips the operator's $T$ parity. For matrix elements that are purely imaginary, we can factor out $i$ from the matrix elements so the bootstrap equations are real-valued.

\subsection{Numerical setup and results}
We begin with a set of primary operators $\mathcal{A}$ and use the equations of motion (\ref{eq:time-derivative-and-matrix-element}) and (\ref{eq:spacial-derivative-and-matrix-element}) to generate descendants up to level $\Lambda$, then take the product of these operators to generate crossing equations of the form (\ref{eq:crossing-equations-general-TFIM}). The matrix elements are further constrained by a number of selection rules (\ref{eq:spin-selection-rule}), (\ref{eq:parity-selection-rule}) and (\ref{eq:phase-selection-rule}). 
In this paper we focus on the single matrix element case where for all crossing equations the external states are the vacuum $\< 0 | \CO \CO' | 0 \>$.
After careful analysis we arrive at the following crossing equations in the block vector form
\begin{equation}\label{eq:crossing-equation-block-form-TFIM}
\begin{aligned}
    0 = & \vec{c}_{0}^T \big(\vec{\mathcal{S}}_0 - \vec{\mathcal{T}}_0 \big) \vec{c}_0
        + \sum_{\Delta E_k, |p|_k} \vec{c}_{k,\uparrow}^T \vec{\mathcal{S}}_{k,\uparrow} (\Delta E_k, |p|_k) \vec{c}_{k,\uparrow} 
        + \sum_{\Delta E_k, |p|_k} \vec{c}_{k,\downarrow}^T \vec{\mathcal{S}}_{k,\downarrow}(\Delta E_k, |p|_k) \vec{c}_{k,\downarrow} 
        \, ,
\end{aligned}
\end{equation}
where $\uparrow$ and $\downarrow$ represent the channels for spin charge $s = +$ and $-$. 
The summation runs over all energy $\Delta E_k \equiv E_k - E_0$ and absolute values of momentum $0\leq |p|_k \leqslant \pi$ of the excited states. 
The $\vec{c}$ vectors represent the set of primary matrix elements (i.e. the ``OPE coefficients'') that are allowed by symmetry. 
Specifically, $\vec{c}_0$ contains all non-vanishing matrix elements between the vacuum state and itself $\< 0 | \CO | 0 \>$, and $\vec{c}_{k,s}$ contains all non-vanishing matrix elements between the vacuum state and the excited states $\< 0 | \CO | E_k, |p|_k, r_k, s \>$, for each spin separately. 
Each $\vec{c}$ can contain a single or multiple primary matrix elements, and the polynomial matrix problems are close to the single and mixed correlator case of the conformal bootstrap. To be concrete, in this work we use the ``single correlator'' setup by taking $\mathcal{A} = \{ \sigma^x, \sigma^z \}$ and ``mixed correlator'' setup by including one more primary operator $\mathcal{A} = \{ \sigma^x, \sigma^z, \sigma_i^x \sigma_{i+1}^z \}$. We leave the study of larger mixing system of primaries and mixing different external states to future work.


Now we can express the crossing equations (\ref{eq:crossing-equation-block-form-TFIM}) as polynomial matrix problems and use semidefinite programming to solve them. We have two setups, corresponding to whether we scan for specific values of the elements in $\vec{c}_0$, which is analogous to ``OPE search'' setup in \cite{Chester:2019ifh}. 

\begin{figure}[htbp]
    \centering
    \includegraphics[width=0.6\linewidth]{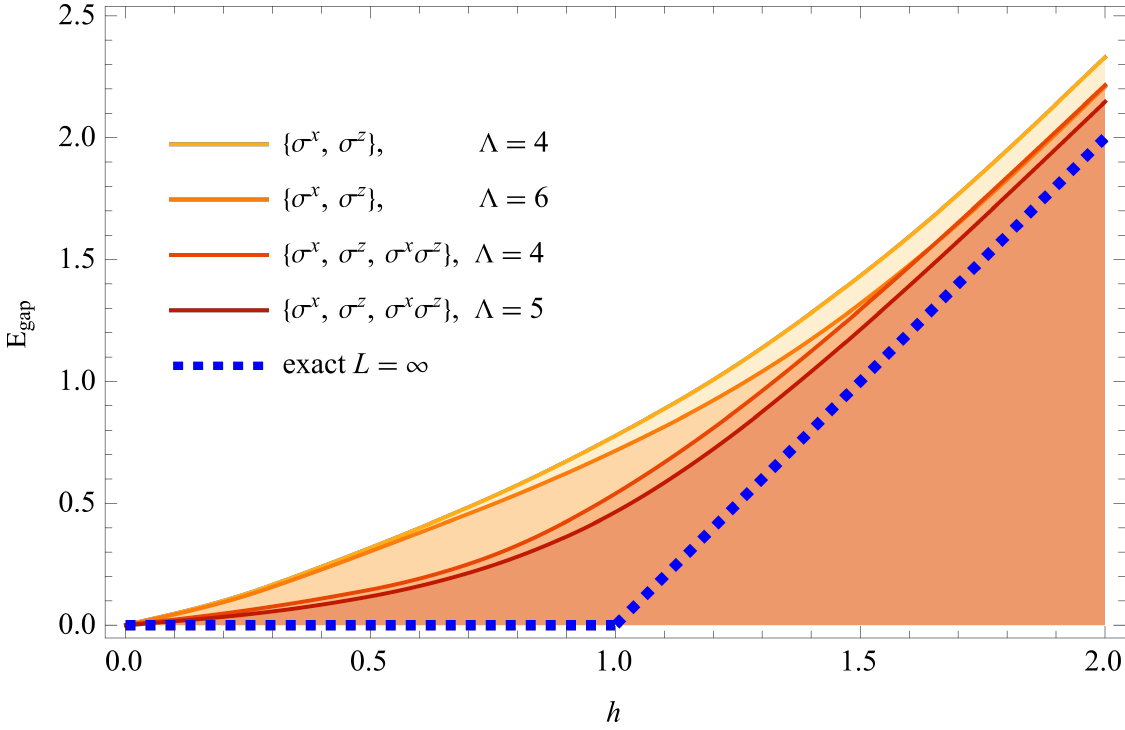}
    \caption{Upper bound on the gap of the $(1+1)$-dimensional transverse field Ising model as a function of field strength $h$ .
    The colors of the lines and shaded regions correspond to bootstrap setups with different choices of the primary operator set $\mathcal{A}$ and truncation level $\Lambda$ on the derivative operators. The bound improves as we include more primary operators and take higher truncation on derivatives. The bound approaches the exact solution faster at $h$ away from the critical point. }
    \label{fig:TFIM-binary}
\end{figure}
First we introduce the basic setup for bootstrapping the gap:
\begin{equation}\label{eq:bootstrap-problem-spectral-TFIM}
\boxed{
\begin{aligned}
&\text{If there exists $\alpha_{I}$ such that $\forall \Delta E_k \geqslant E_{\rm gap}$ and $0 \leqslant |p|_k \leqslant \pi$}\\ 
&\sum_{I} \alpha_{I} 
\big(\vec{\CS}_0 - \vec{\CT}_0\big)_{I} \succ 0 \\ 
& \sum_{I} \alpha_{I} \big(\vec{\CS}_{k,\uparrow} (\Delta E_k, |p|_k) \big)_{I} \succeq 0 \\ 
& \sum_{I} \alpha_{I} \big(\vec{\CS}_{k,\downarrow} (\Delta E_k, |p|_k) \big)_{I} \succeq 0 \, , \\ 
& \text{then all spectra with the prescribed $E_{\rm gap}$ are ruled out.}
\end{aligned}
}
\end{equation}
We use \verb|sdpb| \cite{Simmons-Duffin:2015qma,Landry:2019qug} to efficiently search for the vector $\alpha_I$. Unlike the conformal bootstrap problems, where there is usually only one continuous variable, the scaling dimension $\Delta$, in the spectral bootstrap problem both $p_k$ and $\Delta E_k$ are continuous. Semidefinite programming can be generalized to multivariate polynomial matrix problems with some adaptations in the numerical techniques, and still produces rigorous and optimal bounds on the spectrum\cite{Dyer:2017rul,twoVarBootstrap}. We explain the adaptations and discuss some numerical subtleties we encounter in appendix~\ref{sec:multivariate}. The benchmark performance of these jobs are summarized in Table~\ref{tab:performance}.
We show the numerical bootstrap bounds on the gap in Figure~\ref{fig:TFIM-binary} \footnote{
        The bounds with $\mathcal{A} = \{ \sigma^x, \sigma^z, \sigma_i^x \sigma_{i+1}^z \}$ are obtained using the less rigorous method of discretizing $p$.
        For two variable semidefinite programming there is a rigorous method explained in appendix~\ref{sec:multivariate}. But an alternative approach is to just take a dense discrete set of $p$ and require positivity of the functional only at those values of $p$, and the semidefinite problem is univariate again. The discretization makes it possible for the bound to appear stronger than the true bound because the functional does not need to be positive between the discrete points. To ensure rigorousness one can check the positivity of the functional by hand, but in practice we just run the same problem with higher and higher discretization until the bound no longer changes. We emphasize that the $\mathcal{A} = \{ \sigma^x, \sigma^z \}$ bounds are rigorous because we take the rigorous approach.
    }. 
The bound makes no approximation. For truncation level $\Lambda$ the crossing equations assumes the lattice size $L$ cannot be smaller than $(2\Lambda +1)$. The bootstrap problem (\ref{eq:bootstrap-problem-spectral-TFIM}) assumes no ground state degeneracy. Therefore, we expect that the TFIM spectra for all $L \geqslant 2\Lambda +1$ and $h \geqslant 1$ should obey the bound, and indeed as $\Lambda$ increases the upper bound is stronger as more and more finite volume solutions are ruled out. For $h<1$, even if the assumption of no ground state degeneracy is inconsistent with the theoretical expectation of the broken phase at infinite volume, the finite volume solutions may still be consistent with the assumptions, so the $h<1$ parameter space is not ruled out. The bounds are not saturated by the corresponding $L = 2\Lambda +1$ exact diagonalization solution, suggesting that we have not fully utilized all the self-consistency conditions and that the bounds may improve as more operators or external states are included into the mixing system.

\begin{figure}
    \centering
    \includegraphics[width=0.6\linewidth]{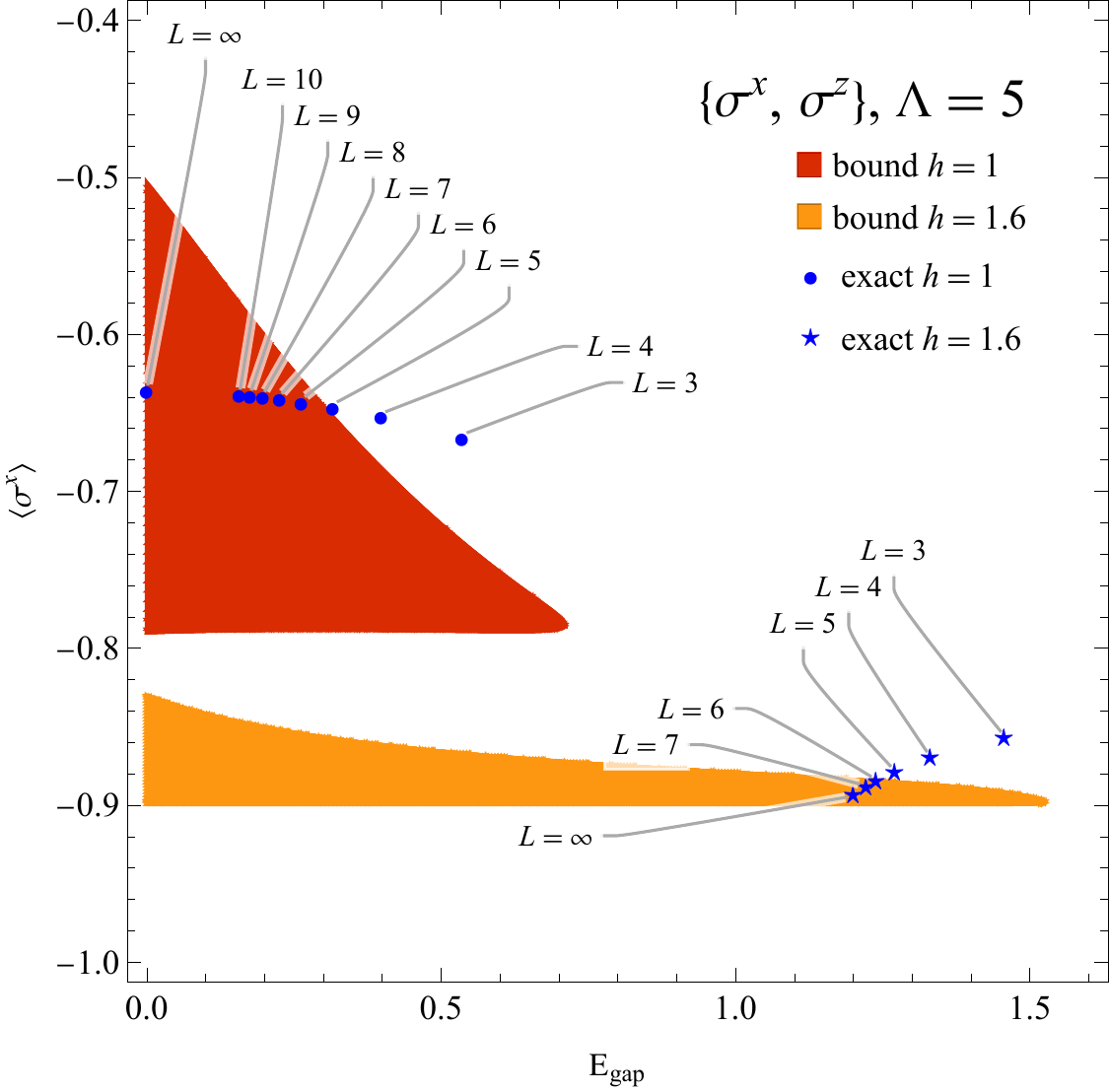}
    \caption{
        The bootstrap bound on the gap and the $\< \sigma^x \>$ vacuum expectation value at $h=1$ (red) and $h=1.6$ (orange). The shaded regions represent the allowed regions from bootstrap while the blue points indicate the exact diagonalization result. The numerical bootstrap setup takes the set of primary operators to be $\mathcal{A} =  \{ \sigma^x, \sigma^z \}$, and the maximum descendant level $\Lambda = 5$. 
    }
    \label{fig:TFIM-c-scan}
\end{figure}
Next we introduce the setup that scans the expectation value $\< \sigma^x \>$ together with the gap. 
With either $\mathcal{A} = \{ \sigma^x, \sigma^z \}$ or $\mathcal{A} = \{ \sigma^x, \sigma^z, \sigma_i^x \sigma_{i+1}^z \}$, the vacuum diagonal matrix element vector only contains $\sigma^x$ due to the spin selection rule, $\vec{c}_0 = \big(\begin{matrix}  1 & \< \sigma^x \>  \end{matrix}\big)$. 
So $\< \sigma^x \>$ will be the only vacuum expectation value considered in this paper. The polynomial matrix problem is:
\begin{equation}\label{eq:bootstrap-problem-spectral-TFIM-ope}
\boxed{
\begin{aligned}
&\text{If there exists $\alpha_{I}$ such that $\forall \Delta E_k \geqslant E_{\rm gap}$ and $0 \leqslant |p|_k \leqslant \pi$}\\ 
&\sum_{I} \alpha_{I} \big(\begin{matrix}  1 & \< \sigma^x \>  \end{matrix}\big)
\big(\vec{\CS}_0 - \vec{\CT}_0\big)_{I} 
\left( \begin{array}{c}
     1  \\
     \< \sigma^x  \>  
\end{array}\right)
= 1 \\ 
& \sum_{I} \alpha_{I} \big(\vec{\CS}_{k,\uparrow} (\Delta E_k, |p|_k) \big)_{I} \succeq 0 \\ 
& \sum_{I} \alpha_{I} \big(\vec{\CS}_{k,\downarrow} (\Delta E_k, |p|_k) \big)_{I} \succeq 0 \, , \\ 
& \text{then all spectra with the prescribed $E_{\rm gap}$ and $\< \sigma^x \>$ are ruled out.}
\end{aligned}
}
\end{equation}
In scanning $\< \sigma^x \>$ we do not implement the OPE scan algorithm in \cite{Chester:2019ifh}. In the future when we have larger system of mixing, the OPE scan will be a useful tool. We show a two-dimension plot of the bound on the gap $E_{\rm gap}$ and the expectation value  $\< \sigma^x \>$ at $h=1$ and $h=1.6$  in Figure~\ref{fig:TFIM-c-scan}. 
We see that the allowed regions are shaped like wedges and the shape and position of the wedges depend on $h$. The allowed region for the critical coupling $h=1$ is still large, suggesting that bounding the critical point is more difficult. At $h=1.6$ the allowed region is significantly smaller. The upper bound on the gap is close to the exact solution, and $\< \sigma^x \>$ is bounded on a narrow band, providing a good estimation of the true value with a rigorous error bar. 

\begin{table}[htbp]
    \centering
    \begin{tabular}{|c|c|c|c|}
        \hline  
        set of primaries $\mathcal{A}$ & $\Lambda$ & number of equations & sdpb time\\
        \hline
        $\{ \sigma^x, \sigma^z\}$ & 4 & 39 & $229s$ \\
        \hline
        $\{ \sigma^x, \sigma^z\}$ & 5 & 56 & $900s$ \\
        \hline
        $\{ \sigma^x, \sigma^z\}$ & 6 & 76 & $1909s$ \\
        \hline
        $\{ \sigma^x, \sigma^z, \sigma_i^x\sigma_{i+1}^z\}^*$ & 4 & 95 & $671s$ \\
        \hline
        $\{ \sigma^x, \sigma^z, \sigma_i^x\sigma_{i+1}^z\}^*$ & 5 & 132 & $1622s$ \\
        \hline
    \end{tabular}
    \caption{Benchmark performance of the spectral bootstrap on the TFIM. The starred setups use the less rigorous method of discretization and are faster, and the other setups use the rigorous two-variable semidefinite programming method. The sdpb time is the total time of a single sdpb job on 4 cpus without hot-starting. The hot-starting can typically speed-up the computation by a factor of 10.  }
    \label{tab:performance}
\end{table}


\section{Discussion}
Our broad goal is to determine the infrared phase of a strongly coupled system from the fundamental description in ultraviolet limit. Monte-Carlo simulation and variational methods are both enormously successful, but these methods both approximate the true physical quantities and introduce statistical or theoretical errors. The bootstrap approach makes no approximation and provides rigorous error bars to the physical quantities, but so far the bootstrap approach is most successful in areas that rely on little or no knowledge of the underlying UV description. The matrix bootstrap approach is a new method that combines the the rigorous nature of bootstrap with knowledge of the UV description. The method uses the UV equations of motion and positivity to impose bounds on the physical quantities.
We have extended this approach to one capable of imposing bounds on the gap of an infinite lattice by constraining the off-diagonal matrix elements. 
Our method uses the consistency between the correlator and its spectral decomposition as well as the Hamiltonian equations of motion to impose bounds on physical quantities including the gap, expectation values, and transition amplitudes. The framework is general and we have tested it on a quantum anhamonic oscillator and the $(1+1)$-dimension transverse field Ising model. In the anharmonic oscillator case, we show that the spectral bootstrap significantly improves the matrix bootstrap method. The lowest two energy eigenvalues and the matrix elements between these states are solved to $10^{-8}$ precision with a minimal number of crossing equations. In the case of the transverse field Ising model, we show that spectral bootstrap imposes a non-trivial bound on the gap and the $\sigma^x$ vacuum expectation value. The bound improves as we increase the number of crossing equations and primary operators in the setup. 

We would like to emphasize several advantages of the spectral bootstrap approach. First of all, this is a method that evaluates the gap and other quantities of an infinite lattice without any approximation, and error bars are rigorous. Second, the spectral bootstrap systematically leverages off-diagonal matrix element information that is absent in the matrix bootstrap, so the bound can be much stronger while considering the same set of operators. 
Finally, the spectral decomposition makes it possible to impose constraints directly on the deep IR physics. For the matrix bootstrap such information is encoded in the expectation value of operators that have a large spatial extent, and is currently inaccessible. In the spectral bootstrap, the information of the deep IR is encoded in the low energy regime of the spectral decomposition, which can be directly accessed by imposing gaps and assumptions in the decomposition. 

The spectral bootstrap is in its infancy and it has considerable room for improvement. The bounds have not converged and it is still an open question whether we can push the bounds to saturation at the thermodynamic limit. The most straightforward next step is to consider a larger number of primary operators in $\mathcal{A}$, as well as mixing with other states. The state mixing setup is similar to the anharmonic oscillator mixed matrix element study, where the bounds improve qualitatively. The current spectral bootstrap does not include any assumption that is specific to the model at infinite volume, and in fact any finite volume solution with $L \geqslant (2\Lambda +2)$ will not be ruled out via bootstrap. Thus it is important to find safe assumptions that can formally rule out the finite volume solutions. An example of such an assumption is that the energy density of all low energy states are the same in the thermodynamic limit. 
The numerical adaptation of semidefinite programming to the two variable polynomial matrix problem is experimental, and it is not at its maximum efficiency. A systematic treatment of the multivariate problem will likely enhance the efficiency of the spectral bootstrap method and will have many more applications in physics, such as the higher-genus modular bootstrap. 
The basis operators do not need to be local operators. The transverse field Ising model is dual to a free fermion theory where the fermion creation operators are highly nonlocal in the original spin basis. It is conceivable that the spectral bootstrap is more powerful for spin systems in a basis of nonlocal operators such as domain wall creation operators or matrix product operators (MPO). For models like the Hubbard model and lattice $\phi^4$ model, it may be advantageous to take the basis operators to be Fourier transforms of local operators. 
We would like to also consider future generalizations and applications of the spectral bootstrap method. The current setup can be immediately applied to lattice $\phi^4$ theory. It is straightforward to consider a generalization to higher spacetime dimensions, with some additional treatment of the discrete rotation symmetry. 
It will also be interesting to consider how to generalize the spectral bootstrap to Quantum Field Theories, where Lorentz symmetry can help further classify the matrix elements but where there is additional challenge from the UV divergence.

\begin{center}
\subsection*{Acknowledgments}
\end{center}
We thank 
Meng Cheng, Rajeev Erramili, Liam Fitzpatrick, Ami Katz, Yuezhou Li, Zhijin Li, Ian Moult, João Penedones, David Poland, Jiaxin Qiao, Balt van Rees, Victor Rodriguez, Junchen Rong, Marco Serone, Ning Su, Xi Yin, and Zechuan Zheng
for useful discussions. We thank Hongbin Chen for collaboration in the early stage of this work. YX is supported by a Mossman Prize Fellowship at Yale University. 






\appendix


\section{Details of anharmonic oscillator crossing equations}\label{sec:mix-details-AHO}
In this appendix we show the details of the mixed matrix element bootstrap crossing equation
\begin{equation}\tag{\ref{eq:mixed-equations-AHO}}
\begin{aligned}
0 &= \sum_{k_-} 
\big( \begin{matrix} c_{0k,x} & c_{1k,x^2} \end{matrix}\big) \vec{\CS}_- 
\left( \begin{matrix} c_{0k,x} \\ c_{1k,x^2} \end{matrix}\right)
+ \sum_{k_+} 
\big( \begin{matrix} c_{0k,x^2} & c_{1k,x} \end{matrix}\big) \vec{\CS}_+ 
\left( \begin{matrix} c_{0k,x^2} \\ c_{1k,x} \end{matrix}\right) \\
&\quad + \big( \begin{matrix} 1 & c_{00,x^2} & c_{11,x^2} & c_{01,x} \end{matrix}\big)
\big( \vec{\CS}_{\rm c} - \vec{\mathcal{T}}_{\rm c} \big)
\big( \begin{matrix} 1 & c_{00,x^2} & c_{11,x^2} & c_{01,x} \end{matrix}\big)^{T} \, .
\end{aligned}
\end{equation}
The vector blocks $\vec{\CS}$ and $\vec{\CT}$ contain what is analogous to the ``conformal blocks'', and in our cases they are the combinations of $g_i^{(\CO)}$ factors. 
We introduce the short-hand notation
\begin{equation}
G_{i,j}^{\<k|\CO|\ell|\CO'|m\>} \equiv g_{i}^{(\CO)}(E_k,E_\ell) g_{j}^{(\CO')}(E_\ell,E_m)
\end{equation}
to reduce the size of the expression. 
First we have an isolated channel $\vec{\CS}_{\rm c}$ and $\vec{\CT}_{\rm c}$ denoting the s-channel and t-channel respectively. The matrix elements involved are the matrix elements between the external states $|0\>$ and $|1\>$, i.e. the matrix elements $\< 0 | \CO | 0 \>$, $\< 1 | \CO | 1 \>$, and $\< 0 | \CO | 1 \>$. The block vector has 3 pieces, $\vec{\CS}_{\rm c}^{\<0ij0\>}$, $\vec{\CS}_{\rm c}^{\<1ij1\>}$ and $\vec{\CS}_{\rm c}^{\<0ij1\>}$ from the  $\< 0 | x^i x^j | 0 \>$, $\< 1 | x^i x^j | 1 \>$ and $\< 0 | x^i x^j | 1 \>$ crossing equations, respectively
\begin{subequations}
\begin{align}
\vec{\CS}_{\rm c}^{\<0ij0\>} &= \left(\begin{matrix}
    G_{i,j}^{\<0|\mathbbm{1}|0|\mathbbm{1}|0\>}
    & G_{i,j}^{\<0|\mathbbm{1}|0|x^2|0\>}
    & 0 & 0
    \\
    G_{i,j}^{\<0|x^2|0|\mathbbm{1}|0\>}
    & G_{i,j}^{\<0|x^2|0|x^2|0\>}
    & 0 & 0
    \\ 
    0 & 0 & 0 & 0 \\ 
    0 & 0 & 0 & 
    G_{i,j}^{\<0|x|1|x|0\>}
\end{matrix}\right) \\ 
\vec{\CT}_{\rm c}^{\<0ij0\>} &= \left(\begin{matrix}
    g_{i+j}^{(\mathbbm{1})}(E_0) 
    & \frac{1}{2} g_{i+j}^{(x^2)}(E_0) 
    & 0 & 0
    \\
    \frac{1}{2} g_{i+j}^{(x^2)}(E_0) 
    & 0
    & 0 & 0
    \\ 
    0 & 0 & 0 & 0 \\ 
    0 & 0 & 0 & 
    0
\end{matrix}\right) \\ 
\vec{\CS}_{\rm c}^{\<1ij1\>} &= \left(\begin{matrix}
    G_{i,j}^{\<1|\mathbbm{1}|1|\mathbbm{1}|1\>}
    & 0
    & G_{i,j}^{\<1|\mathbbm{1}|1|x^2|1\>}
    & 0 
    \\
    0 & 0 & 0 & 0 \\ 
    G_{i,j}^{\<1|x^2|1|\mathbbm{1}|1\>}
    & 0 
    & G_{i,j}^{\<1|x^2|1|x^2|1\>}
    & 0
    \\ 
    0 & 0 & 0 & 
    G_{i,j}^{\<1|x|0|x|1\>}
\end{matrix}\right) \\ 
\vec{\CT}_{\rm c}^{\<1ij1\>} &= \left(\begin{matrix}
    g_{i+j}^{(\mathbbm{1})}(E_1) 
    & 0
    & \frac{1}{2} g_{i+j}^{(x^2)}(E_1) 
    & 0 
    \\
    0 & 0 & 0 & 0 \\ 
    \frac{1}{2} g_{i+j}^{(x^2)}(E_1) 
    & 0 
    & 0
    & 0
    \\ 
    0 & 0 & 0 & 0
\end{matrix}\right) \\ 
\vec{\CS}_{\rm c}^{\<0ij1\>} &= \frac{1}{2}\left(\begin{matrix}
    0 & 0 & 0 & 
    G_{i,j}^{\<0|\mathbbm{1}|0|x|1\>} + G_{i,j}^{\<0|x|1|\mathbbm{1}|1\>}  \\ 
    0 & 0 & 0 & 
    G_{i,j}^{\<0|x^2|0|x|1\>} \\ 
    0 & 0 & 0 & 
    G_{i,j}^{\<0|x|1|x^2|1\>} \\ 
    G_{i,j}^{\<0|\mathbbm{1}|0|x|1\>} + G_{i,j}^{\<0|x|1|\mathbbm{1}|1\>} & G_{i,j}^{\<0|x^2|0|x|1\>} & G_{i,j}^{\<0|x|1|x^2|1\>} & 0
\end{matrix}\right) \\
\vec{\CT}_{\rm c}^{\<0ij1\>} &= \frac{1}{2}\left(\begin{matrix}
    0 & 0 & 0 & 
    g_{i+j}^{(x)}(E_0,E_1)  \\ 
    0 & 0 & 0 & 0 \\ 
    0 & 0 & 0 & 0 \\ 
    g_{i+j}^{(x)}(E_0,E_1) & 0 & 0 & 0
\end{matrix}\right) \, .
\end{align}
\end{subequations}
Similarly, we have the even parity channel
\begin{subequations}
\begin{align}
\vec{\CS}_{+}^{\<0ij0\>} &= \left(\begin{matrix}
    G_{i,j}^{\<0|x^2|k|x^2|0\>} 
    & 0
    \\
    0 & 0
\end{matrix}\right) \\ 
\vec{\CS}_{+}^{\<1ij1\>} &= \left(\begin{matrix}
    0 & 0
    \\
    0 & G_{i,j}^{\<1|x|k|x|1\>} 
\end{matrix}\right) \\ 
\vec{\CS}_{+}^{\<0ij1\>} &= \left(\begin{matrix}
    0 & \frac{1}{2}G_{i,j}^{\<0|x^2|k|x|1\>} 
    \\
    \frac{1}{2}G_{i,j}^{\<0|x^2|k|x|1\>}  & 0
\end{matrix}\right) \, .
\end{align}
\end{subequations}
Finally we have the odd-parity channel
\begin{subequations}
\begin{align}
\vec{\CS}_{-}^{\<0ij0\>} &= \left(\begin{matrix}
    G_{i,j}^{\<0|x|k|x|0\>} 
    & 0
    \\
    0 & 0
\end{matrix}\right) \\ 
\vec{\CS}_{-}^{\<1ij1\>} &= \left(\begin{matrix}
    0 & 0
    \\
    0 & G_{i,j}^{\<1|x^2|k|x^2|1\>} 
\end{matrix}\right) \\ 
\vec{\CS}_{-}^{\<0ij1\>} &= \left(\begin{matrix}
    0 & \frac{1}{2}G_{i,j}^{\<0|x|1|x^2|1\>} 
    \\
    \frac{1}{2}G_{i,j}^{\<0|x|1|x^2|1\>}  & 0
\end{matrix}\right) \, .
\end{align}
\end{subequations}

\section{Multivariate semidefinite programming and numerical subtleties}\label{sec:multivariate}
\subsection{Map positive multivariate polynomials to sum of squares}
In this subsection we discuss how to solve multivariate bootstrap problems using semidefinite programming rigorously. The procedure is similar to \cite{Simmons-Duffin:2015qma} with some differences caused by adding more variables. We will focus on the two-variable case for simplicity but it is straightforward to generalize to any number of variables.
We define a two-variable polynomial matrix problem
\begin{equation}\label{eq:pmp}
\begin{aligned}
    &\text{Maximize } b\cdot \alpha \text{ over } \alpha \in \mathbb{R}^N, \\
    &\text{such that } M_j^0(x,y) + \sum_{n=1}^N \alpha_n M_j^n(x,y) \succeq 0 \text{ for all } x,y\in \mathbb{R} \text{ and } 1\leqslant j \leqslant J \\
    &\text{where } M_j^n \equiv \left( \begin{matrix}
        P_{j,11}^n(x,y) & \cdots & P_{j,1m_j}^n(x,y) \\
        \vdots & \ddots & \vdots \\
        P_{j,m_j1}^n(x,y) & \cdots & P_{j,m_jm_j}^n(x,y) \\
    \end{matrix}\right)
\end{aligned}
\end{equation}
Whether semidefinite programming can provide a rigorous and optimal solution requires that for any non-negative polynomial $P(x,y)$ is a sum of squares of polynomials. Equivalently, there is a positive semidefinite matrix $Y$ such that
\begin{equation}\label{eq:sos}
    P(x,y) = {\rm Tr}_{\mathbb{R}^{\delta}} (Y Q_{\delta}(x,y) )
\end{equation}
where $Q$ is made up of a dimension-$\delta$ vector of basis polynomials in $(x,y)$.
\begin{equation}
\begin{aligned}
    Q_\delta (x,y) &\equiv \vec q_\delta (x,y) \vec q_\delta (x,y)^T\\
    \vec q_\delta (x,y) &\equiv \big( q_1 (x,y), q_2 (x,y) \cdots q_{\delta} (x,y) \big)
\end{aligned}
\end{equation}
The terms and dimension of $q_\delta$ depend on the requirement that $Q_\delta$ must contain all degrees of freedom in $P(x,y)$. The requirement is satisfied for a univariate polynomial $P(x)$. For a multivariate polynomial $P(x,y)$ this is no longer true, and a non-negative polynomial is a sum of square of rational function instead, but \cite{reznick1995uniform} shows that for any strictly positive polynomial $P(x,y)$, $(1+x^2+y^2)^g P(x,y)$ satisfies (\ref{eq:sos}) at sufficiently high $g$. Thus our strategy is to run semidefinite programming with polynomials $(1+x^2+y^2)^g P(x,y)$ at $g=0$ first, and we may get sub-optimal bound as some positive functionals may be out of reach if $g$ is insufficient. Then we can increase $g$ and the bound will monotonically improve until all positive functionals are accessible by a sum of squares, and the bound is optimal. In practice, in the spectral bootstrap we never have to take $g$ to be nonzero at all. 
The rest of the problem is book-keeping. Assuming the right hand side of (\ref{eq:pmp}) can be expressed as a sum of square, we have the ansatz 
\begin{equation}\label{pmp-sos}
    P_{j,rs}^0 (x,y) + \sum_n \alpha_n P_{j,rs}^n (x,y) = {\rm Tr}_{\mathbb{R}^{\delta}} \big(Y (Q_{\delta}(x,y)\otimes E^{rs}) \big) \, ,
\end{equation}
where $(E^{rs})_{ij} = \delta_i^r\delta_j^s + \delta_j^r\delta_i^s$ is the element symmetric matrices. The polynomial equality (\ref{pmp-sos}) for all $x,y$ is equivalent to the equality at sufficiently many sample points $\{ (x_k,y_k) \}$,
\begin{equation}
\begin{aligned}
    &P_{j,rs}^0 (x_k,y_k) + \sum_n \alpha_n P_{j,rs}^n (x_k,y_k) = {\rm Tr}_{\mathbb{R}^{\delta}} \big(Y (Q_{\delta}(x_k,y_k)\otimes E^{rs}) \big) \\
    &\forall k = 1,2 \cdots d_j
\end{aligned}
\end{equation}
where the dimension $d_j$ is set by the number of all possible monomials below the leading powers of $P_{j,rs}^n(x,y)$. Now we have translated the polynomial matrix problem input into a numerical equation. Finally, we map the problem to a standard semidefinite programming problem (SDP)
\begin{equation}
\begin{aligned}
    &\text{Maximize } b\cdot \alpha \text{ over } \alpha \in \mathbb{R}^N ,~ Y \in \mathcal{S}^{K}, \\
    &\text{such that } {\rm Tr}(A_p Y) + B_p \cdot \alpha = c_p \text{ and } \\
    &Y \succeq 0
\end{aligned}
\end{equation}
where $\mathcal{S}^{K}$ is the space of $K\times K$ symmetric matrices. The matrices in the above SDP are
\begin{equation}
\begin{aligned}
    Y &= {\rm diag}( Y_1, Y_2, \cdots , Y_J) \quad Y_j \in \mathcal{S}^{m_j\delta_j} \\
    A_p &= {\rm diag}( 0,0 \cdots, Q_{\delta_j}\otimes E^{rs}, \cdots , 0) \quad p \equiv (j,r,s,k)\\
    B_p &= -P_{j,rs}^n(x_k,y_k) \\
    c_p &= P_{j,rs}^0(x_k,y_k) \, .
\end{aligned}
\end{equation}
In the bootstrap algorithm, we bypass \verb|pvm2sdp| and use Mathematica code to directly generate these matrices $A, B$ and $c$ as the SDP input. The internal code for \verb|sdpb| also needs to be slightly changed to allow more flexible values of $\delta_j$ and $d_j$. 

To implement the problems (\ref{eq:bootstrap-problem-spectral-TFIM}) and (\ref{eq:bootstrap-problem-spectral-TFIM-ope}) we take the change of variables $\Delta E = E_{\rm gap} + x^2$ and $e^{ip} = \frac{1-\xi^2}{1+\xi^2} + i \frac{2\xi}{1+\xi^2}$, such that the space of $\Delta E \geqslant E_{\rm gap}$ and $-\pi \leqslant p\leqslant \pi$ are mapped to $x,y \in \mathbb{R}$.

\subsection{Subtleties in the numerical study}

The mixed setup taking $\mathcal{A} = \{ \sigma^x, \sigma^z, \sigma_i^x\sigma_{i+1}^z \}$ has a subtlety that eventually leads us to use discretization instead of the rigorous method. We find that when we have block forms $\vec\CS(x,y)$ that is a vector of $n\times n$ matrices $n>1$, \verb|sdpb| fails to find a positive functional $\vec\alpha \cdot \vec\CS(x,y)\succeq0$. For simplicity, we consider univariate $\vec\CS(x)$. When the highest powers do not match, for example
\begin{equation}\label{eq:mismatching-power}
    \vec\alpha \cdot \vec\CS(x) \sim \left(\begin{matrix}
        a x^n & b x^n \\
        b x^n & c x^{n-2}
    \end{matrix}\right) \sim x^n \left(\begin{matrix}
        a & b \\
        b & 0
    \end{matrix}\right)\, ,
\end{equation}
the functional is manifestly non-positive, and is negative unless $\vec\alpha \cdot \vec\CS(x)$ realizes $b=0$ exactly. This means the functional will be exactly positive semidefinite. SDPB only scans the space of positive definite matrices, and exactly semidefinite matrices can only be reached as an approximation that has a large feasibility error. Empirically, \verb|--detectDualFeasibilityJump| will never work. The solution is to shift the diagonal matrix elements in $\vec\CS(x)$ with an infinitestimal $\epsilon$-shift by $ \epsilon x^{\rm degree}$ in order to push the solution away from being positive semidefinite to being slightly positive definite. While there exists a rigorous approach which is to scan for the mismatching powers and to eliminate $b$ in (\ref{eq:mismatching-power}) before running sdpb, the quick-and-dirty $\epsilon$-shift is faster and still provides an accurate bound. Nevertheless, in the two variable case, the $\epsilon$-shift does not address the issue anymore. We conjecture that the problem is related to the fact that $\vec\CS(x)$ has more than one leading powers, making it tricky how to shift away all the non-positivity. Addressing this subtlety is beyond the scope of this paper, so we resort to discretization in the mixed setup.

\bibliographystyle{JHEP}
\bibliography{refs}

\end{document}